\documentclass{aastex62}

\submitjournal{ApJ}

\begin{document}

\title{Gravitationally Lensed Quasar SDSS J1442+4055: Redshifts of Lensing Galaxies, Time 
Delay, Microlensing Variability, and Intervening Metal System at $z \sim$ 2}
\shorttitle{Gravitational Lens System SDSS J1442+4055}

\correspondingauthor{Vyacheslav N. Shalyapin}
\email{vshal@ukr.net, goicol@unican.es}

\author{Vyacheslav N. Shalyapin}
\affil{O.Ya. Usikov Institute for Radiophysics and Electronics \\
National Academy of Sciences of Ukraine \\
12 Acad. Proscury St., 61085 Kharkiv, Ukraine}
\affil{Departamento de F\'\i sica Moderna \\
Universidad de Cantabria \\
Avda. de Los Castros s/n, 39005 Santander, Spain}

\author{Luis J. Goicoechea}
\affil{Departamento de F\'\i sica Moderna \\
Universidad de Cantabria \\
Avda. de Los Castros s/n, 39005 Santander, Spain}



\begin{abstract}
We present an $r$--band photometric monitoring of the two images A and B of the 
gravitationally lensed quasar SDSS J1442+4055 using the Liverpool Telescope (LT). From the 
LT light curves between 2015 December and 2018 August, we derive at once a time delay of 
25.0 $\pm$ 1.5 days (1$\sigma$ confidence interval; A is leading) and microlensing 
magnification gradients below 10$^{-4}$ mag day$^{-1}$. The delay interval is not expected 
to be affected by an appreciable microlensing--induced bias, so it can be used to estimate 
cosmological parameters. This paper also focuses on new Gran Telescopio Canarias (GTC) and 
LT spectroscopic observations of the lens system. We determine the redshift of two bright 
galaxies around the doubly imaged quasar using LT spectroscopy, while GTC data lead to 
low--noise individual spectra of A, B, and the main lensing galaxy G1. The G1 spectral 
shape is accurately matched to an early--type galaxy template at $z$ = 0.284, and it has 
potential for further relevant studies. Additionally, the quasar spectra show absorption by 
metal--rich gas at $z \sim$ 2. This dusty absorber is responsible for an extinction bump at 
a rest--frame wavelength of 2209 $\pm$ 2 \AA, which has strengths of $\sim$ 0.47 and 0.76 
mag $\mu$m$^{-1}$ for A and B, respectively. In such intervening system, the dust--to--gas 
ratio, gas--phase metallicity indicator [Zn/H], and dust depletion level [Fe/Zn] are 
relatively high. 
\end{abstract}



\section{Introduction} \label{sec:intro}

Gravitationally lensed quasars are becoming essential tools to study the structure and 
composition of galaxies at different redshifts, and of the entire Universe 
\citep{2006glsw.conf.....M,2010ARA&A..48...87T}. Hence, significant effort is being devoted 
to the discovery of lensed quasars and to their follow--up observations. For example, the 
current archive of the Sloan Digital Sky Survey \citep[SDSS;][]{2000AJ....120.1579Y} 
includes photometric and spectroscopic data of more than 500000 quasars 
\citep{2018A&A...613A..51P}. The SDSS database contains a large collection of quasar 
spectra taken as part of the Baryon Oscillation Spectroscopic Survey 
\citep[BOSS;][]{2013AJ....145...10D}, and \citet{2016MNRAS.456.1595M} took advantage of 
this fact to find 13 new double quasars. Other ongoing projects are also reporting 
discoveries of double/quadruple quasars and lists of lensed quasar candidates 
\citep[e.g.,][]{2018MNRAS.480.5017A,2018MNRAS.476..663K,2018MNRAS.479.5060L}. In addition, 
new multiple quasars must be fully characterised by follow--up observations. Detailed 
spectroscopy is used to identify intervening objects, analyse their gas, dust and stellar 
content, and put constraints on the size and structure of sources through 
microlensing--induced spectral distortions \citep[e.g.,][]{2003A&A...405..445W,
2007A&A...468..885S,2011ApJ...730...16M,2016A&A...596A..77G}. Light curves of lensed 
quasars are also key pieces to determine time delays and constrain cosmological parameters 
\citep[e.g.,][]{2008A&A...488..481V,2017MNRAS.465.4914B,2017ApJ...836...14S}, and/or detect 
microlensing variability, and thus learn about the structure of quasars and the composition 
of lensing galaxies \citep[e.g.,][]{2004ApJ...605...58K,2011A&A...528A.100S,
2013ApJ...774...69H}.

After performing data mining to identify double quasar candidates in the SDSS-III DR10 
\citep{2014ApJS..211...17A,2014A&A...563A..54P}, complementary observations of some of such 
candidates allowed \citet{2016MNRAS.456.1948S} to find the new optically bright, wide 
separation double quasar \object{SDSS J1442+4055} \citep[see also][]{2016MNRAS.456.1595M}. 
From a medium--resolution SDSS--BOSS spectrum of the A image of \object{SDSS J1442+4055}, 
\citet{2014A&A...563A..54P} presented several estimates of the source (quasar) redshift. 
While the SDSS pipeline produced $z_{\rm{s}}$ = 2.5746 $\pm$ 0.0002, a principal component 
analysis led to $z_{\rm{s}}$ = 2.5931 $\pm$ 0.0006. After a visual inspection, 
\citet{2014A&A...563A..54P} indicated that $z_{\rm{s}}$ = 2.593, and we adopt this value 
throughout the paper. The lens system basically consists of two quasar images (A and B) 
having $r \sim$ 18$-$19 mag and separated by 2\farcs156, the main lensing galaxy (G1; $r 
\sim$ 19.5 mag) located 1\farcs38 from the A image, and two additional intervening objects: 
a secondary galaxy (G2; $r \sim$ 19.8 mag) in the vicinity of G1 and an absorber at $z$ = 
1.946 \citep{2016MNRAS.456.1948S}. 

This paper is dedicated to describe and deeply analyse follow--up observations of 
\object{SDSS J1442+4055}. In the framework of the Gravitational LENses and DArk MAtter 
(GLENDAMA) project \citep{2018A&A...616A.118G}, in Section \ref{sec:data}, we present 
a 2.7--year photometric monitoring with the 2.0 m Liverpool Telescope (LT) in the $r$ band 
and associated light curves for both quasar images, as well as spectroscopic observations 
with the 10.4 m Gran Telescopio Canarias (GTC) and the LT. The high signal--to--noise ratio 
(SNR) spectroscopy over a wide wavelength interval with the GTC allows us to accurately 
identify the primary lensing galaxy G1, while we use the LT spectroscopic data of two 
bright secondary galaxies (G2 and another object in the field around the quasar) to measure 
their redshifts. In Section \ref{sec:delay}, the cross--correlation between the LT light 
curves of A and B yields the time delay in the lens system. The $r$--band microlensing 
variability is also 
discussed in Section \ref{sec:delay}. In Section \ref{sec:dust}, using the GTC spectra of A 
and B, we reveal the origin of the flux ratios $B/A$ from near UV to near IR. In Section 
\ref{sec:gas}, we study in detail the intervening gas at $z \sim$ 2 and its connection with
dust at the same redshift. In addition to the GTC and SDSS--BOSS spectra, this analysis is 
also based on data from the MMT Observatory, i.e., medium--resolution UV--visible spectra 
(blueward of 6000 \AA) of the two quasar images with moderate SNR 
\citep{2018ApJS..236...44F}. Our main results and conclusions are summarised in Section 
\ref{sec:final}.  

When we were completing our third monitoring season and this work, 
\citet{2018A&A...619A.142K} presented and analysed Keck spectra of \object{SDSS 
J1442+4055}. In this paper, their spectroscopic results are compared to ours.

\section{Observations and data reduction} \label{sec:data}

\subsection{LT--IO:O photometric monitoring} \label{subsec:ioo}

We have monitored the double quasar \object{SDSS J1442+4055} over three complete observing 
seasons from 2015 December 23 to 2018 August 28. This monitoring suffers from two 
visibility gaps of about four months, which are not important in determining the relatively 
short time delay between images (see Section \ref{sec:delay}). All photometric observations 
were made in the $r$ band with the IO:O camera on the LT. With 2$\times$2 binning (pixel 
scale of 0\farcs30), we obtained frames of the lensed quasar for 135 observing epochs 
(nights). For the first seven nights, we took two consecutive exposures of 250 s each, 
while 4$\times$150 s exposures were obtained on every remaining night. Thus, we collected 
526 individual frames of 150 or 250 s. The LT pipeline performed a primary data reduction 
including bias subtraction, overscan trimming, and flat fielding. Additionally, we 
interpolated over bad pixels and removed cosmic rays.

\begin{figure}[h!]
\centering
\includegraphics[width=0.7\textwidth]{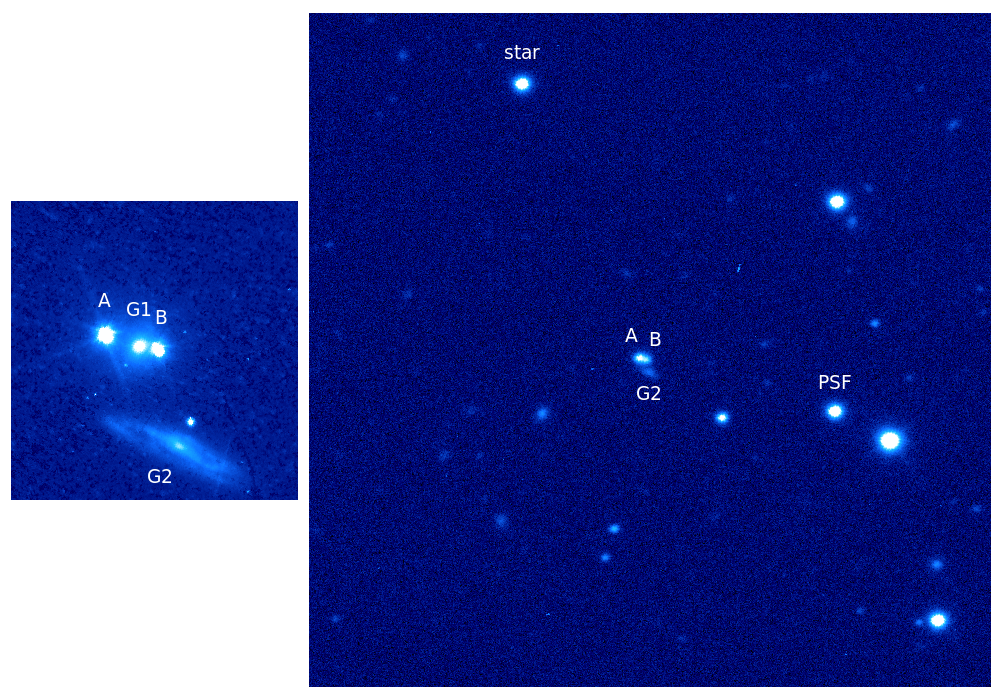}
\caption{$HST$ and LT imaging of SDSS J1442+4055. Left: $HST$--WFC3--G280 
zeroth--order frame of size $10\arcsec \times 10\arcsec$. Right: First LT--IO:O $r$--band 
frame on 2015 December 22. This covers a field of view of $200\arcsec \times 200\arcsec$, 
and includes the PSF star and the control star (see main text).}
\label{fig:f1} 
\end{figure}

In order to extract fluxes for the two quasar images, we used the IMFITFITS software 
\citep{1998AJ....115.1377M}. From this tool, assuming reasonable brightness profiles for 
the lensing galaxies G1 and G2, one can determine the fluxes of A and B through 
point--spread function (PSF) fitting. The brightness of the main deflector G1 was modelled 
as a de Vaucouleurs profile \citep{2016MNRAS.456.1595M}, whereas we taken an exponential 
profile to model the light distribution of the secondary lens G2, since the {\it Hubble 
Space Telescope} ($HST$) archive includes an image of G2 showing the presence of a disc 
and spiral arms\footnote{Program Id: 14127, PI: Michele Fumagalli} (see the left panel 
of Figure~\ref{fig:f1}). Both profiles were then convolved with the empirical PSF from the 
close star at RA (J2000) = 220\fdg706485 and Dec. (J2000) = +40\fdg922076 ($r$ = 16.075 
mag; see the right panel of Figure~\ref{fig:f1}). This PSF star was also used to 
model the point--like sources A and B. The relative positions of B and G1 (with respect to 
A), as well as the ellipticity, orientation, and effective radius of G1, were taken from 
\citet{2016MNRAS.456.1948S}. In addition, the ellipticity and orientation of G2 were set to  
their values in the SDSS database. In a first iteration, the fluxes of G1 and G2, and the
relative position and effective radius of G2, were estimated from the best frames in terms 
of SNR and FWHM seeing. In a second iteration, we applied the code to all individual 
frames, allowing the position of A, the background level, and the fluxes of A and B to be 
free. We also calculated PSF fluxes for a field (control) star that is located at RA 
(J2000) = 220\fdg741396 and Dec. (J2000) = +40\fdg949647 ($r$ = 15.998 mag; see the 
right panel of Figure~\ref{fig:f1}). 

\begin{deluxetable}{ccccccc}[h!]
\tablecaption{LT--IO:O $r$--band light curves of SDSS J1442+4055AB.\label{tab:t1}}
\tablenum{1}
\tablewidth{0pt}
\tablehead{
\colhead{MJD--50000} & 
\colhead{$m_{\rm{A}}$\tablenotemark{a}} & \colhead{$em_{\rm{A}}$\tablenotemark{a}} & 
\colhead{$m_{\rm{B}}$\tablenotemark{a}} & \colhead{$em_{\rm{B}}$\tablenotemark{a}} & 
\colhead{$m_{\rm{S}}$\tablenotemark{ab}} & \colhead{$em_{\rm{S}}$\tablenotemark{ab}}
}
\startdata
7379.288 & 18.076 & 0.005 & 19.008 & 0.009 & 15.998 & 0.004 \\
7398.293 & 18.029 & 0.006 & 19.012 & 0.012 & 15.993 & 0.006 \\
7400.289 & 18.022 & 0.006 & 19.018 & 0.011 & 16.000 & 0.005 \\
7406.279 & 18.026 & 0.005 & 18.998 & 0.009 & 15.993 & 0.005 \\
7412.267 & 18.034 & 0.007 & 18.982 & 0.012 & 15.994 & 0.006 \\
\enddata
\tablenotetext{a}{$r$--SDSS magnitude.}
\tablenotetext{b}{We use S to denote the field (control) star.}
\tablecomments{Table 1 is published in its entirety in the machine-readable format.
A portion is shown here for guidance regarding its form and content.}
\end{deluxetable}

\begin{figure}[h!]
\centering
\includegraphics[width=0.7\textwidth]{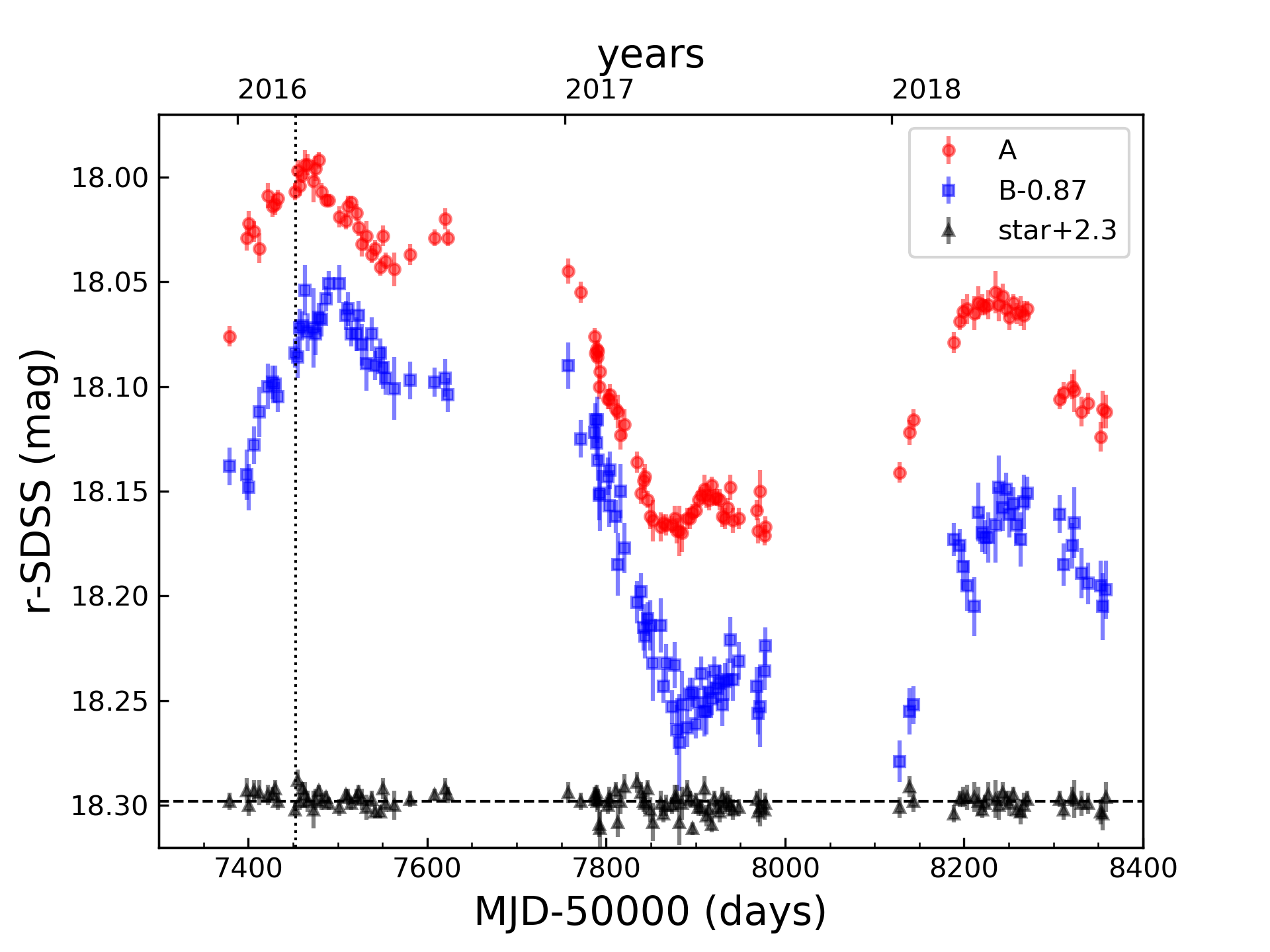}
\caption{LT--IO:O $r$--band light curves of A, B, and the control star. The curves of B and 
the star are shifted by $-$0.87 and $+$2.3 mag, respectively, to facilitate comparison. The
vertical dotted line corresponds to the epoch of our spectroscopic observations with 
GTC--OSIRIS (see Section \ref{subsec:osiris}), while the horizontal dashed line represents 
the constant flux of the control star.} 
\label{fig:f2}
\end{figure}

The quasar light curves ($r$--SDSS magnitudes) show anomalous results for 51 individual 
frames. These frames producing outliers are characterised by high FWHM seeing, low SNR for 
A or tracking/guiding errors (very elongated or trailed stars), and thus, we removed them 
from the final database. The remaining 475 frames represent 90\% of the individual 
observations and have a median FWHM of 1\farcs28. We then combined magnitudes measured on 
the same night to obtain final photometric data at 135 epochs. To estimate typical 
photometric errors in the light curves of A, B, and the control star, we used deviations 
between magnitudes having time separations $\leq$ 3 days. This statistical analysis led to 
uncertainties of 0.0051 (A), 0.0093 (B), and 0.0046 (star) mag, which were multiplied by 
the relative SNR at each epoch, $\left< \rm{SNR} \right>$/SNR, to calculate errors on a 
nightly basis \citep[$\left< \rm{SNR} \right>$ is the average SNR;][]{howell06}. Our final 
light curves of A, B, and the star are available in Table \ref{tab:t1} and shown in 
Figure~\ref{fig:f2}. 

\subsection{GTC--OSIRIS spectroscopy} \label{subsec:osiris}

We performed spectroscopic observations of \object{SDSS J1442+4055} on 2016 March 5 using 
the OSIRIS instrument on the GTC. We took a 2850 
(3$\times$950) s GTC--OSIRIS exposure with each of the two grisms R500B and R500R, and used 
IRAF\footnote{IRAF is distributed by the National Optical Astronomy Observatory, which is 
operated by the Association of Universities for Research in Astronomy (AURA) under 
cooperative agreement with the National Science Foundation. This software is available at 
\url{http://iraf.noao.edu/}} packages to carry out data reductions. All 950 s 
sub--exposures were obtained in dark time, at low airmasses and under good seeing 
conditions. The average values of the airmass and the FWHM seeing at 6225 \AA\ amounted to 
1.03 and 0\farcs89, respectively. Regarding the dispersions, they were close to the 
nominal ones: $D$(R500B) = 3.560 \AA\ pix$^{-1}$ and $D$(R500R) = 4.814 \AA\ pix$^{-1}$. 
However, we slightly modified the standard wavelength ranges, decreasing the minimum 
wavelength for R500B (3425 \AA; to include relevant absorption features) and the maximum 
wavelength for R500R (9255 \AA; to avoid fringing and second--order contamination). The 
spatial pixel scale was 0\farcs254.

In order to extract spectra of all individual sources in the strong lensing region, the 
1\farcs23--width slit was oriented along the line joining A and B, and we followed a 
technique similar to those in our previous analyses of GTC--OSIRIS spectroscopic data 
\citep[e.g.,][]{2016A&A...596A..77G}. We modelled the lens system as a 2D light
distribution consisting of two point--like sources (A and B) and a circular de Vaucouleurs 
profile with $r_{\rm{eff}}$ = 0\farcs59 (G1), whose relative positions are given in Table 
2 of \citet{2016MNRAS.456.1948S}. Such ideal model was then convolved with a 2D Moffat PSF 
having a power index $\beta$ = 3, masked with the slit transmission and integrated across 
the slit. Apart from the position of A and the FWHM value, our 1D model at each wavelength 
bin 
included the fluxes of A, B, and G1 as free parameters, and thus, fits to the GTC--OSIRIS 1D 
data allowed us to obtain the spectra of the two quasar images and the main lensing galaxy.
For each source, in addition to wavelength--dependent fluxes $F_{\lambda}$, we estimated 
flux errors $eF_{\lambda}$ using the equation (9) of \citet{1986PASP...98..609H}. This 
method for extracting individual spectra is significantly different from the technique used
by \citet{2018A&A...619A.142K}, who considered Keck--LRIS observations in the wavelength 
range 3600--8650 \AA\ on 2016 June 5, extracted data of A and B that are contaminated by 
light from G1, and then fitted templates for the intrinsic spectral slopes of A, B, and G1, 
and the supposed reddening of A and B arising from dust in the absorber at $z$ = 1.946 (we 
justify this hypothesis in Section \ref{sec:dust}).  

\begin{deluxetable}{ccccccc}[h!]
\tablecaption{GTC--OSIRIS--R500B spectra of SDSS J1442+4055ABG1.\label{tab:t2}}
\tablenum{2}
\tablewidth{0pt}
\tablehead{
\colhead{$\lambda$\tablenotemark{a}} & 
\colhead{$F_{\lambda}$(A)\tablenotemark{b}} & 
\colhead{$eF_{\lambda}$(A)\tablenotemark{c}} & 
\colhead{$F_{\lambda}$(B)\tablenotemark{b}} &
\colhead{$eF_{\lambda}$(B)\tablenotemark{c}} & 
\colhead{$F_{\lambda}$(G1)\tablenotemark{b}} &
\colhead{$eF_{\lambda}$(G1)\tablenotemark{c}}  
}
\startdata
3426.770 & 17.928 & 1.660 & 14.001 & 1.520 & -10.045 & 1.694 \\
3430.330 & 19.053 & 1.569 &  9.429 & 1.446 &  -3.607 & 1.645 \\
3433.891 & 19.969 & 1.490 &  8.143 & 1.324 &   0.734 & 1.571 \\
3437.452 & 17.532 & 1.420 &  9.126 & 1.259 &  -3.928 & 1.449 \\
3441.013 & 21.708 & 1.435 &  7.852 & 1.201 &  -2.417 & 1.424 \\
\enddata
\tablenotetext{a}{Observed wavelength in \AA.}
\tablenotetext{b}{Flux in 10$^{-17}$ erg cm$^{-2}$ s$^{-1}$ \AA$^{-1}$.}
\tablenotetext{c}{Flux error in 10$^{-17}$ erg cm$^{-2}$ s$^{-1}$ \AA$^{-1}$.}
\tablecomments{Table 2 is published in its entirety in the machine-readable format.
A portion is shown here for guidance regarding its form and content.}
\end{deluxetable}

\begin{deluxetable}{ccccccc}[h!]
\tablecaption{GTC--OSIRIS--R500R spectra of SDSS J1442+4055ABG1.\label{tab:t3}}
\tablenum{3}
\tablewidth{0pt}
\tablehead{
\colhead{$\lambda$\tablenotemark{a}} & 
\colhead{$F_{\lambda}$(A)\tablenotemark{b}} & 
\colhead{$eF_{\lambda}$(A)\tablenotemark{c}} & 
\colhead{$F_{\lambda}$(B)\tablenotemark{b}} &
\colhead{$eF_{\lambda}$(B)\tablenotemark{c}} & 
\colhead{$F_{\lambda}$(G1)\tablenotemark{b}} &
\colhead{$eF_{\lambda}$(G1)\tablenotemark{c}}  
}
\startdata
4846.734 & 22.387 & 0.362 &  9.399 & 0.274 & 4.743 & 0.305 \\
4851.550 & 22.515 & 0.358 &  9.669 & 0.271 & 4.009 & 0.300 \\
4856.365 & 23.775 & 0.353 &  9.587 & 0.265 & 4.339 & 0.293 \\
4861.180 & 24.849 & 0.346 & 10.300 & 0.261 & 4.242 & 0.287 \\
4865.995 & 23.896 & 0.322 & 10.484 & 0.243 & 2.466 & 0.260 \\
\enddata
\tablenotetext{a}{Observed wavelength in \AA.}
\tablenotetext{b}{Flux in 10$^{-17}$ erg cm$^{-2}$ s$^{-1}$ \AA$^{-1}$.}
\tablenotetext{c}{Flux error in 10$^{-17}$ erg cm$^{-2}$ s$^{-1}$ \AA$^{-1}$.}
\tablecomments{Table 3 is published in its entirety in the machine-readable format.
A portion is shown here for guidance regarding its form and content.}
\end{deluxetable}

\begin{figure}[h!]
	\begin{minipage}[h]{0.5\linewidth}
      \centering
      \includegraphics[width=1.0\textwidth]{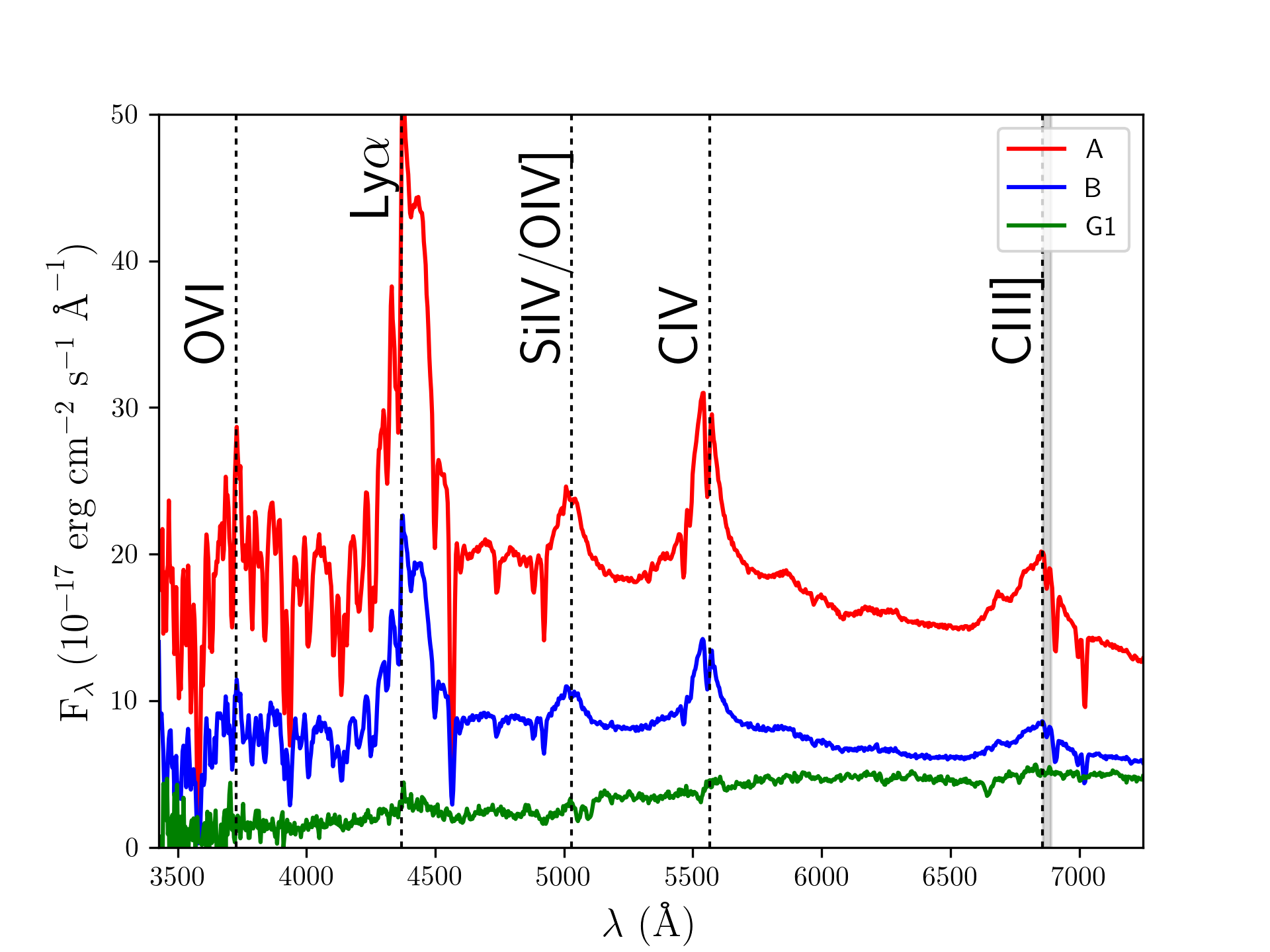}
      \end{minipage}
	\begin{minipage}[h]{0.5\linewidth}
      \centering
      \includegraphics[width=1.0\textwidth]{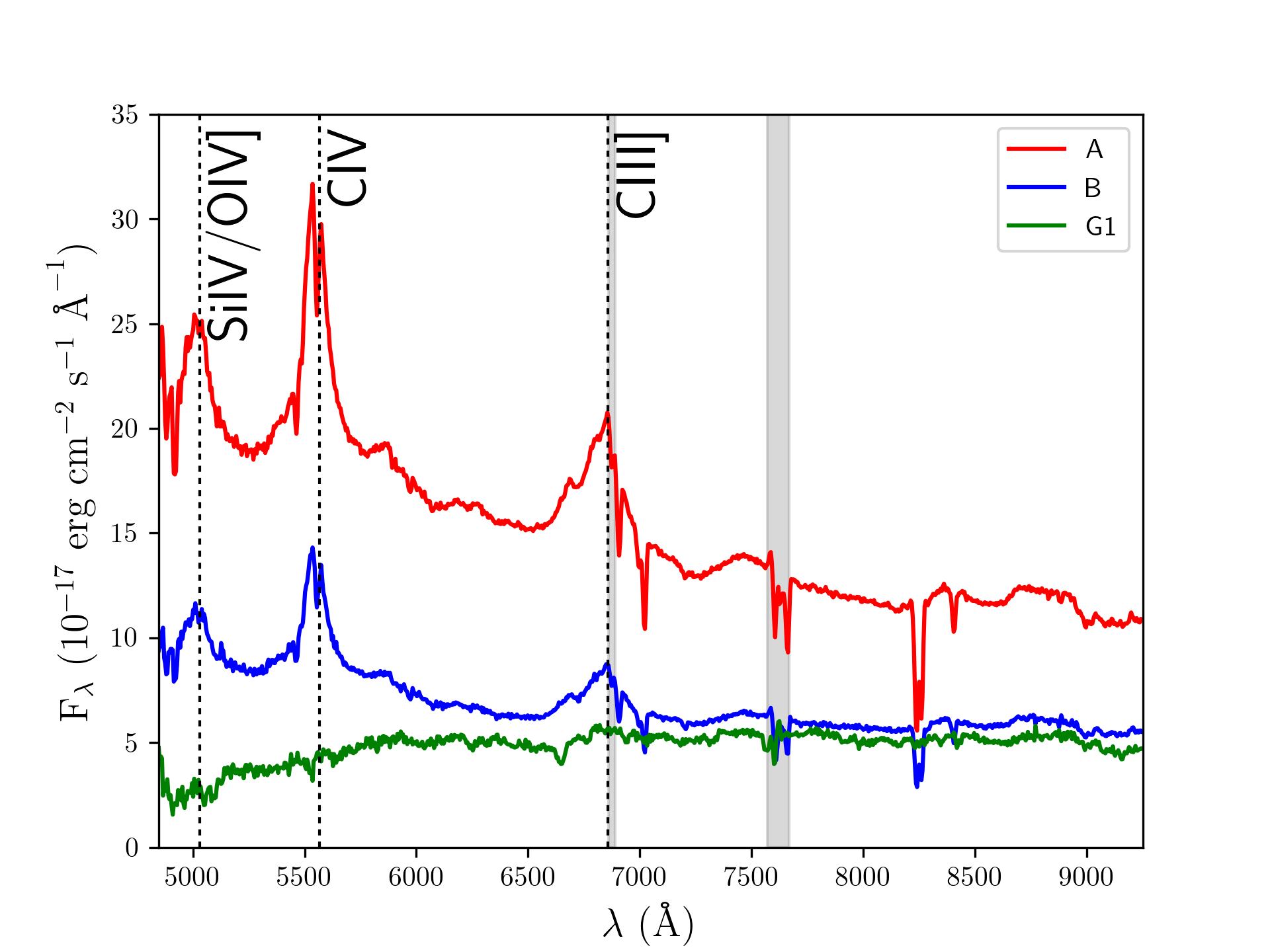}
    	\end{minipage}

\vspace{3.00mm}

	\begin{minipage}[h]{0.5\linewidth}
      \centering
      \includegraphics[width=1.0\textwidth]{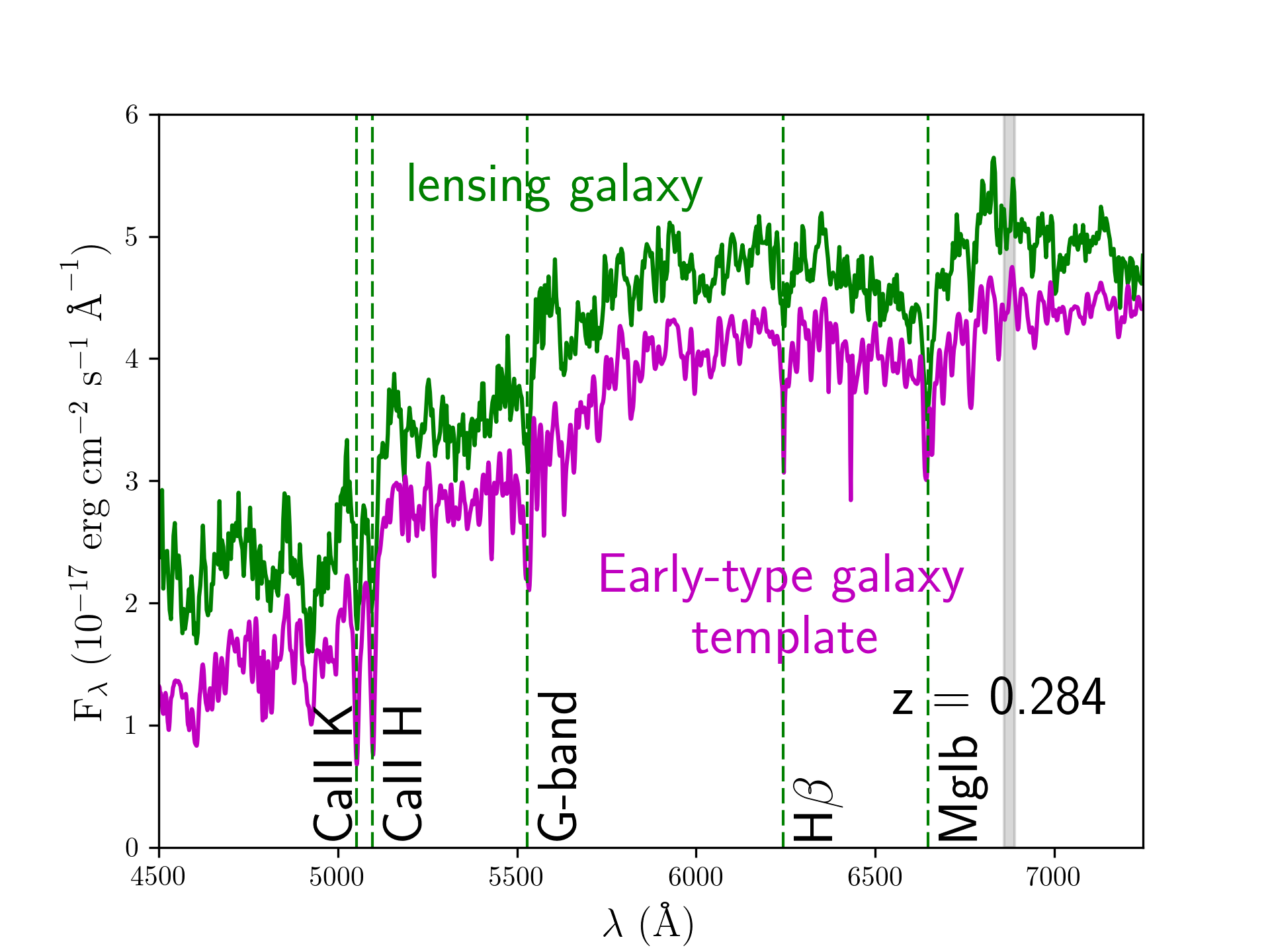}
      \end{minipage}
	\begin{minipage}[h]{0.5\linewidth}
      \centering
      \includegraphics[width=1.0\textwidth]{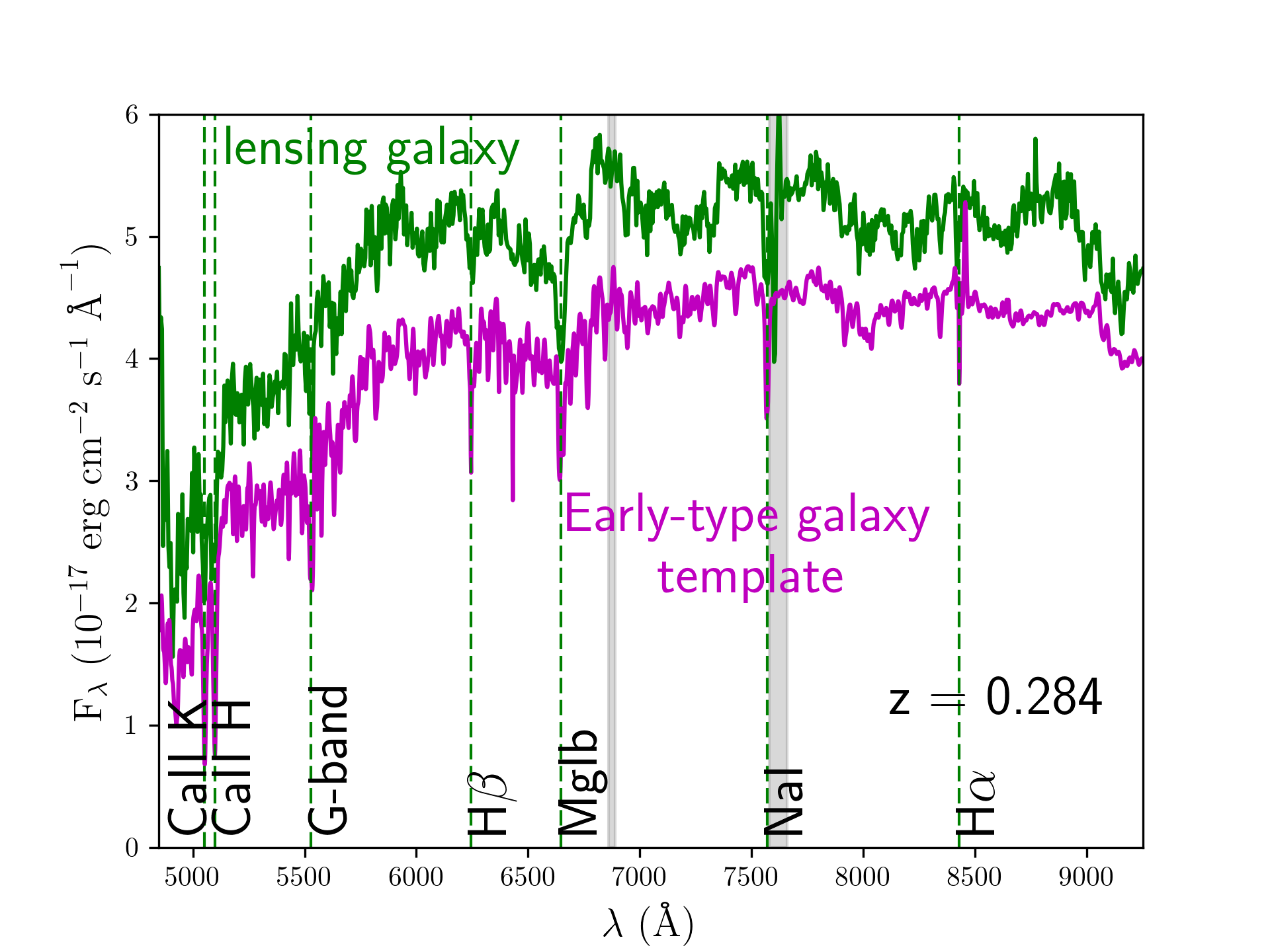}
    	\end{minipage}
\caption{GTC--OSIRIS spectra of SDSS J1442+4055ABG1 in 2016 March. Top: R500B (left) and 
R500R (right) spectra of the three sources in the strong lensing region. Vertical dotted 
lines indicate emission lines at $z_{\rm{s}}$ = 2.593, while grey highlighted bars are 
associated with atmospheric artefacts. Bottom: R500B (left) and R500R (right) spectra of 
the primary lensing galaxy. These two panels display zoomed--in versions of the G1 spectra 
along with the red--shifted ($z$ = 0.284) SDSS spectral template of an early--type galaxy. 
Vertical dashed lines are associated with absorption features.}
\label{fig:f3} 
\end{figure}

We also checked our wavelength and flux calibrations, which were based on HgAr and Ne arc 
lamp exposures, as well as spectra of the standard star Hilt600 \citep{1992PASP..104..533H,
1994PASP..106..566H}. First, we compared positions of narrow absorption lines in the 
GTC--OSIRIS spectra of A and positions of such lines in the SDSS--BOSS spectrum of the 
brightest quasar image, taken on 2012 May 29. This comparison allowed detection of systematic 
deviations in our 
wavelength zero--points, so the R500B and R500R data were shifted by +1.5 \AA\ and $-$2.0 
\AA, respectively. Second, we used $r$--band frames taken with the LT on 2016 March 4 to 
measure $r$--band fluxes of A and B, and compare them to
the corresponding GTC--OSIRIS fluxes. These spectral fluxes agreed well with the LT 
photometry (typical deviation of $\sim$1\%), so the spectral energy distributions were not 
rescaled. The final calibrated spectra of A, B, and G1 are included in Tables \ref{tab:t2} 
and \ref{tab:t3}. In addition, the results from the observations with both grisms are 
plotted in Figure~\ref{fig:f3}. It is also worth mentioning that the spectral shapes of A and
B in the wavelength range 3500--6000 \AA\ are consistent with those obtained with the $HST$
WFC3--G280 grism on 2016 April 21, i.e., about seven weeks 
later \citep{2018ApJ...860...41L}. All raw and reduced frames in FITS format are publicly 
available at the GTC archive\footnote{\url{http://gtc.sdc.cab.inta-csic.es/gtc/index.jsp}}. 

In the top panels of Figure~\ref{fig:f3}, we show the R500B (left) and R500R (right) 
spectra of A (red), B (blue), and G1 (green). The accurate quasar spectra contain five 
prominent emission features at $z_{\rm{s}}$ = 2.593: O\,{\sc vi}, Ly$\alpha$, Si\,{\sc 
iv}/O\,{\sc iv}], C\,{\sc iv}, and C\,{\sc iii}] (vertical dotted lines), and enable us to 
probe flux ratios $B/A$ over a very broad interval of wavelengths from near UV to near IR 
(3430 to 9250 \AA). Furthermore, the GTC--OSIRIS spectra of A and B include an intervening 
metal system (IMS), which was also detected in the SDSS--BOSS spectrum of A 
\citep{2016MNRAS.456.1948S} and the Keck spectra of both quasar images 
\citep{2018A&A...619A.142K}. We measured $z_{\rm{IMS}}$ = 1.9465 using strong Fe\,{\sc 
ii}/Mg\,{\sc ii} absorption lines in the SDSS--BOSS spectral energy distribution. Physical 
properties of this high--$z$ galaxy halo are widely discussed in Sections \ref{sec:dust} 
and \ref{sec:gas}. We do not pay special attention to other absorbers. For example, there 
is a proximate system at $z$ = 2.586 $\sim$ $z_{\rm{s}}$, consisting of neutral hydrogen 
(Ly$\alpha$ and Ly$\beta$ lines) and high--ionisation metals (O\,{\sc vi}, N\,{\sc v}, 
Si\,{\sc iv}, and C\,{\sc iv} lines). This is most probably associated with the quasar host 
galaxy or its environment \citep[e.g.,][]{2010MNRAS.406.1435E}. There are also prominent 
Ly$\alpha$ 
systems at $z$ = 2.578, 2.406, and 2.296, so the $HST$--WFC3--G280 spectra blueward of 3300 
\AA\ are strongly absorbed by neutral hydrogen. The Ly$\alpha$ break at $\lambda$ = 3000 
\AA\ \citep{2018ApJ...860...41L} is mainly due to the nearest Ly$\alpha$ system at $z$ 
= 2.296.   

In the bottom panels of Figure~\ref{fig:f3}, we can appreciate details of the R500B (left) 
and R500R (right) spectra of the lensing galaxy G1 (green). The spectral shapes and the 
positions of several absorption features (vertical dashed lines; e.g., the Ca\,{\sc ii} 
$HK$ doublet, G--band, H$\beta$ line, and Mg\,{\sc i} $b$ triplet) match very well with an 
early--type galaxy template\footnote{SDSS spectral template No. 23 at 
\url{http://classic.sdss.org/dr7/algorithms/spectemplates/index.html}} at $z$ = 0.284 
(magenta). Hence, $z_{\rm{G1}}$ = 0.284 $\pm$ 0.001 was inferred from the positions of the
absorption lines. The G1 spectra and the templates in the bottom panels of 
Figure~\ref{fig:f3} are of a similar quality, whereas the Keck--LRIS spectrum of G1 in Fig. 
3 of \citet{2018A&A...619A.142K} is much more noiser. In any case, our $z_{\rm{G1}}$ value 
fully agrees with the lens redshift from Keck--LRIS data, which was based on a complex fit 
(see above). This consistency of results through different data sets and analysis 
techniques strengthens reliability of the measured redshift.

\subsection{LT--SPRAT data} \label{subsec:sprat}

In the vicinity of the double quasar, there is a secondary lensing galaxy (G2) that is 
displayed in Fig. 4 of \citet{2016MNRAS.456.1948S}. The two galaxies G1 and G2 are $<$ 
5\arcsec\ apart, so they could be physically associated. Indeed, these sources have similar 
$r$--band brightness, and $z_{\rm{G1}}$ (see Section \ref{subsec:osiris}) is consistent 
with the SDSS photometric redshift of G2: 0.323 $\pm$ 0.051. To identify G2 and another 
bright field galaxy (G3), both objects were spectroscopically observed on 2016 June 8. The 
SDSS position of G3 is RA (J2000) = 220\fdg73908 and Dec (J2000) = +40\fdg92188 
(southeast of the quasar images), and thus, A and G3 are separated by 33\farcs9. Moreover, 
G3 is brighter than G2 ($r$ = 19.1) and has a photometric redshift of 0.188 $\pm$ 0.029.

\begin{deluxetable}{ccc}[h!]
\tablecaption{LT--SPRAT spectra of SDSS J1442+4055G2G3.\label{tab:t4}}
\tablenum{4}
\tablewidth{0pt}
\tablehead{
\colhead{$\lambda$\tablenotemark{a}} & 
\colhead{$F_{\lambda}$(G2)\tablenotemark{b}} &
\colhead{$F_{\lambda}$(G3)\tablenotemark{b}}  
}
\startdata
3984.359 &  6.368 & -5.175 \\
3988.991 & 12.435 &  4.964 \\
3993.623 &  3.922 &  3.153 \\
3998.255 &  4.809 &  2.718 \\
4002.887 & -6.239 &  4.339 \\
\enddata
\tablenotetext{a}{Observed wavelength in \AA.}
\tablenotetext{b}{Flux in 10$^{-17}$ erg cm$^{-2}$ s$^{-1}$ \AA$^{-1}$.}
\tablecomments{Table 4 is published in its entirety in the machine-readable format.
A portion is shown here for guidance regarding its form and content.}
\end{deluxetable}
 
\begin{figure}[h!]
\centering
\includegraphics[width=0.7\textwidth]{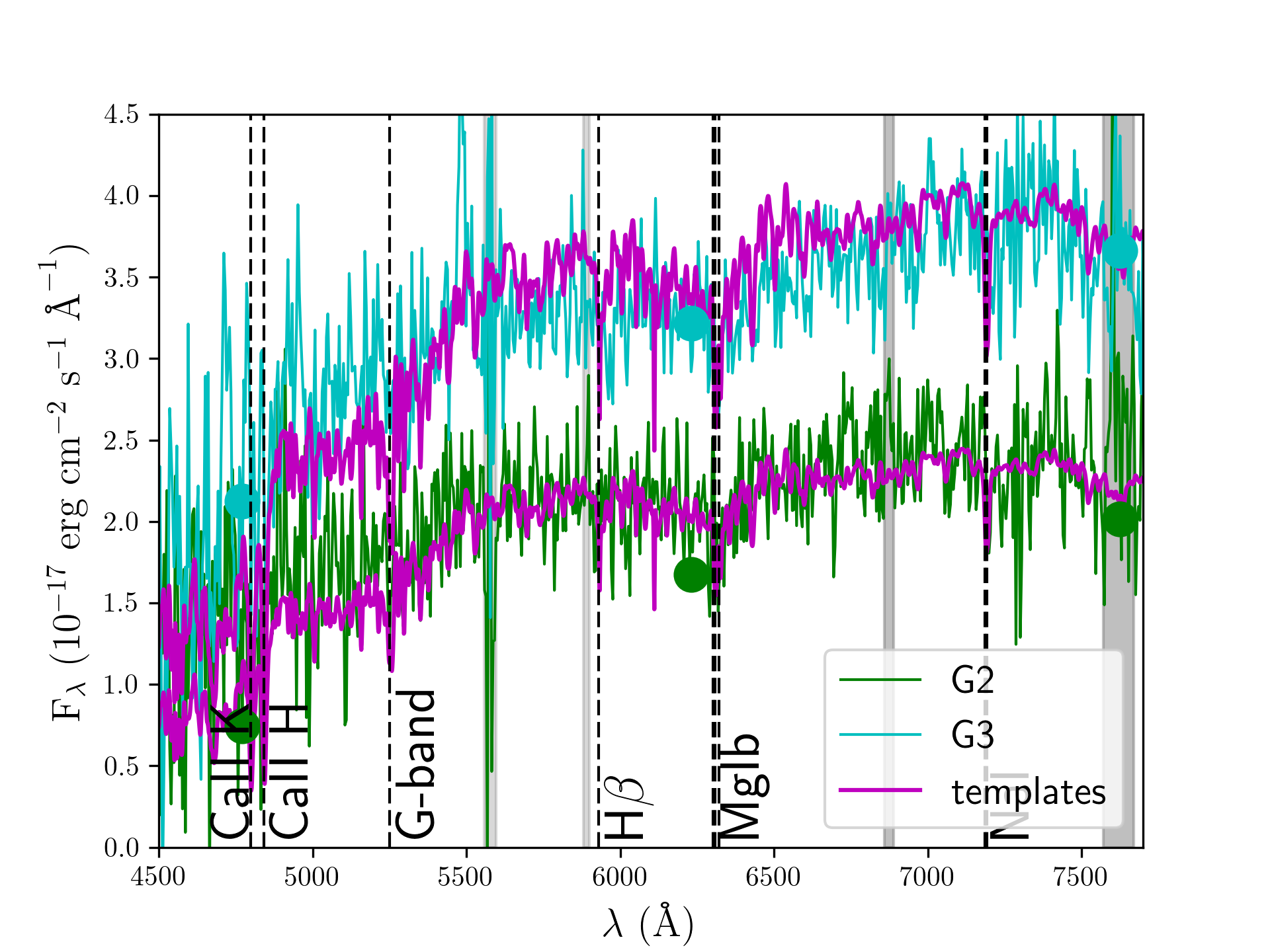}
\caption{LT--SPRAT spectra of SDSS J1442+4055G2G3 in 2016 June. The spectral shapes of G2 
(green) and G3 (cyan) are compared to early--type galaxy templates at $z$ = 0.22 (magenta). 
The green and cyan circles correspond to half of the $gri$ fluxes in the SDSS database 
(see main text). Vertical dashed lines indicate absorption features at $z$ = 0.22, while 
grey highlighted bars are associated with atmospheric artefacts.} 
\label{fig:f4}
\end{figure}

We used the red grating mode of the SPRAT instrument on the LT, which is optimized for the 
red region of the 4000--8000 \AA\ wavelength range. Additionally, the 1\farcs8--width slit 
was oriented along the line joining G2 and G3. The dispersion and spatial pixel scale were 
4.63 \AA\ pix$^{-1}$ and 0\farcs44. We took 5 $\times$ 600 s science exposures under good 
observing conditions: moonless night, FWHM $\sim$ 1\arcsec, and airmass of $\sim$1.1. 
Tungsten lamp and Xe arc exposures were used for flat fielding and wavelength 
calibration, respectively. We also observed the spectrophotometric standard star BD+33d2642 
\citep{1990AJ.....99.1621O} for flux calibration. After a primary reduction of frames under 
the IRAF working environment, the spectra of G2 and G3 were extracted using the task APALL. 
These spectroscopic data are included in Table \ref{tab:t4} and Figure~\ref{fig:f4}. 

In Figure~\ref{fig:f4}, we show the two spectra (green and cyan) and two red--shifted SDSS 
templates of an early--type galaxy (magenta; see Section \ref{subsec:osiris}), along with 
$gri$ fluxes for both galaxies from the SDSS database (circles). Despite G2 is a spiral 
galaxy (see Section \ref{subsec:ioo}), we do not detect any emission line in its visible 
spectrum. The original SDSS fluxes of G2 and G3 were reduced by 50\% to roughly account for 
slit losses, since the slit width does not cover their entire luminous halo. Although the 
LT--SPRAT spectra are quite noisy, their shapes indicated that the two secondary galaxies 
are at similar redshift $z_{\rm{G2}}$ = $z_{\rm{G3}}$ = 0.22 $\pm$ 0.01. Hence, the 
photometric redshift of G2 does not correspond to the true value of $z_{\rm{G2}}$, and this 
galaxy is not physically associated with the main deflector. 

\section{Time delay and microlensing variability} \label{sec:delay}

\begin{figure}[h!]
	\begin{minipage}[h]{1.0\linewidth}
	\centering
	\includegraphics[width=0.7\textwidth]{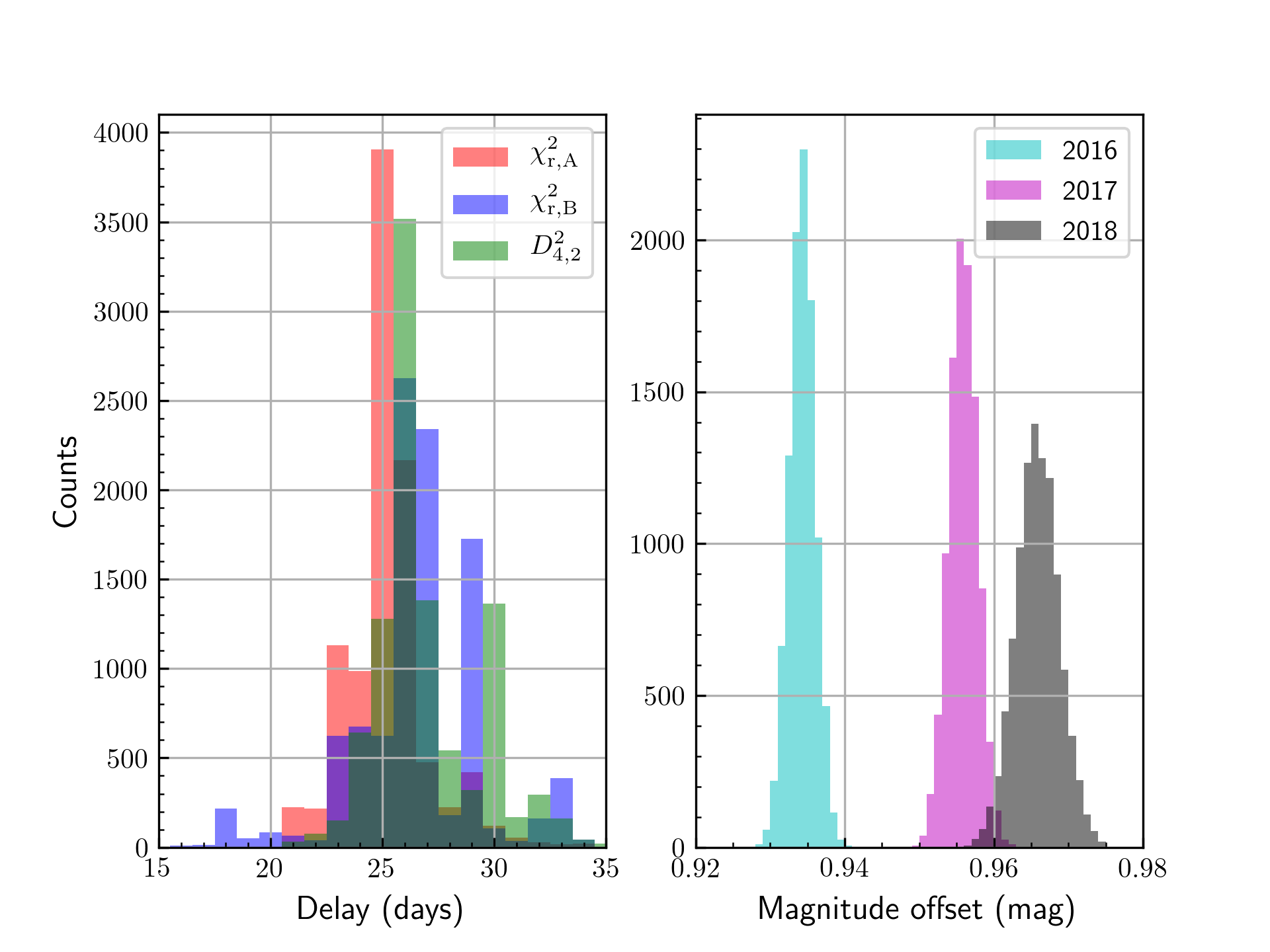}
     	\end{minipage}

\vspace{3.00mm}

	\begin{minipage}[h]{1.0\linewidth}
      \centering
      \includegraphics[width=0.7\textwidth]{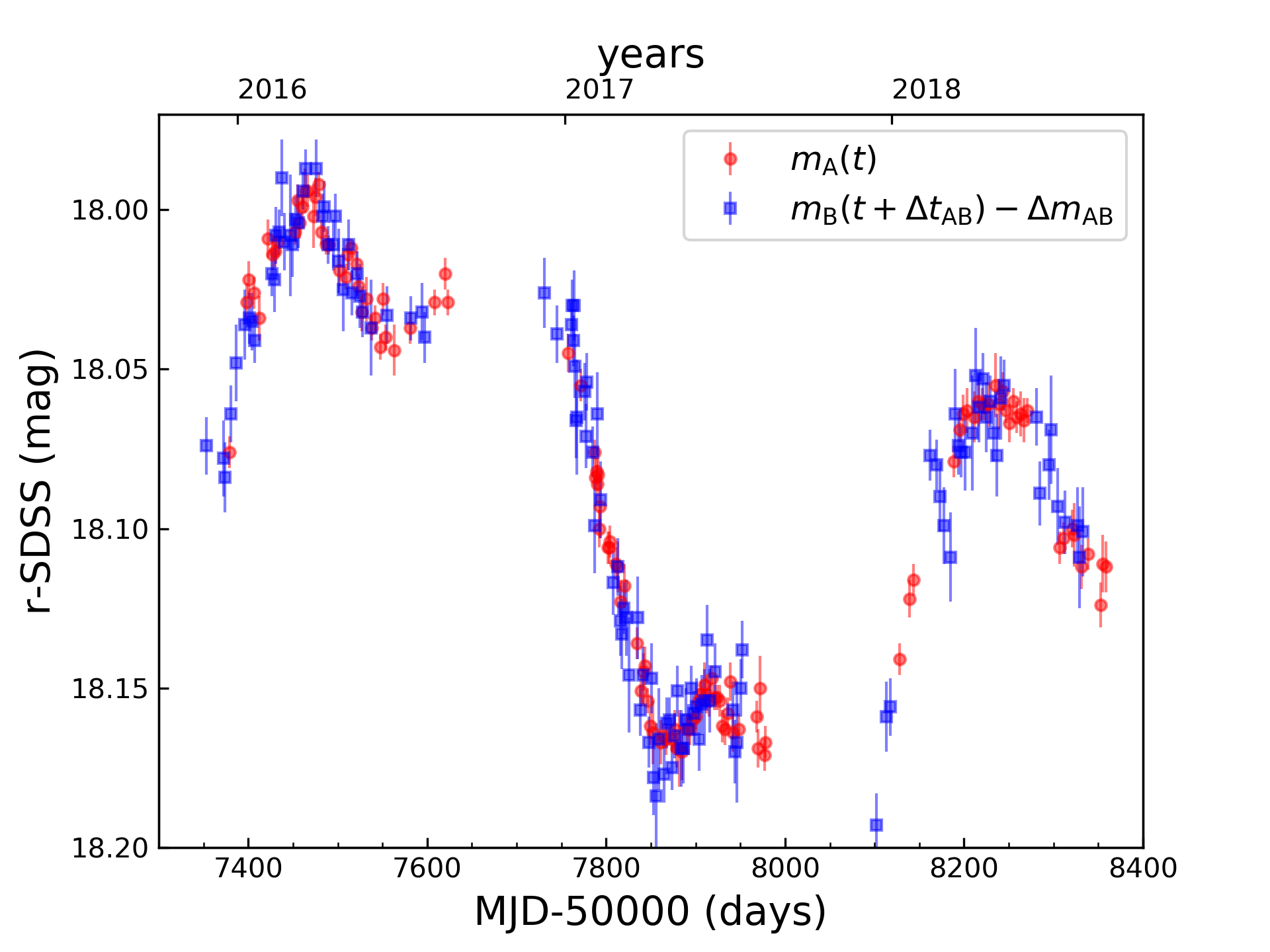}
      \end{minipage}
\caption{Top: Histograms from 10$^4$ pairs of simulated curves AB. The left panel shows the 
best solutions of the time delay from the $D^2_{4,2}$, $\chi^2_{\rm{r,A}}$, and 
$\chi^2_{\rm{r,B}}$ estimators including seasonal microlensing ($\delta = \alpha$ = 10 
days). The right panel displays the best solutions of the magnitude offsets in 2016, 2017, 
and 2018 from the $\chi^2_{\rm{r,A}}$ minimization. Bottom: Combined light curve in the $r$ 
band. The A curve is compared with the magnitude-- and time--shifted B curve ($\Delta 
t_{\rm{AB}}$ = 25 days, $\Delta m_{\rm{AB}}$(2016) = 0.934 mag, $\Delta m_{\rm{AB}}$(2017) 
= 0.956 mag, and $\Delta m_{\rm{AB}}$(2018) = 0.966 mag; see main text).}
\label{fig:f5} 
\end{figure}

The quasar light curves in Figure~\ref{fig:f2} display almost parallel prominent 
variations, which suggest a short time delay between images and a slow microlensing signal.
In this section, we use two standard techniques to measure the time delay, identify the 
microlensing variability, and thus confirm our qualitative conclusion. First, we considered
the dispersion method to match both light curves. More precisely, the $D^2_{4,2}$ estimator 
\citep{1996A&A...305...97P} including a step function--like (seasonal) microlensing. The 
value of the decorrelation length ($\delta$) has little influence on the best solutions for
the time delay ($\Delta t_{\rm{AB}} = \tau_{\rm{B}} - \tau_{\rm{A}}$) and the three 
magnitude offsets (one per season; $\Delta m_{\rm{AB}} = m_{\rm{B}}(t + \Delta t_{\rm{AB}}) 
- m_{\rm{A}}(t)$), and after checking results for 4 $\leq \delta \leq$ 20 days, we chose an 
intermediate value of 10 days to estimate confidence intervals. We generated 10$^4$ 
simulated light curves of each quasar image at epochs equal to those of observation, 
modifying the observed magnitudes by adding random quantities (repetitions of the LT 
experiment). These additive random numbers were realisations of normal distributions around 
zero, with standard deviations equal to the measured uncertainties. The $D^2_{4,2}$ 
estimator ($\delta$ = 10 days) with a step function--like microlensing was then applied to 
each pair (A and B) of simulated curves to produce distributions of delays and magnitude 
offsets. The delay histogram is shown in the top left panel of Figure~\ref{fig:f5}.

Second, we carried out a reduced chi--square ($\chi^2_{\rm{r}}$) minimization with three 
magnitude offsets, i.e., considering a seasonal microlensing similar to that of the 
dispersion method. The $\chi^2_{\rm{r}}$ technique has two variants 
\citep[e.g.,][]{2006A&A...452...25U}: $\chi^2_{\rm{r,A}}$ compares the curve A with the 
time--shifted and binned curve B, and $\chi^2_{\rm{r,B}}$ compares the curve B with the 
time--shifted and binned curve A. In both variants, bins are characterised by a semisize 
$\alpha$, which plays a role similar to the decorrelation length in $D^2_{4,2}$. Reasonable
values of $\alpha$ (in the interval 4$-$20 days) led to similar best solutions for the 
delay and the magnitude offsets, so we focused on results for $\alpha$ = 10 days. It is 
also worth mentioning that the best solutions for $\alpha$ = 10 days correspond to 
$\chi^2_{\rm{r,A}}$ = 0.69 and $\chi^2_{\rm{r,B}}$ = 0.91. This means that the seasonal
microlensing scenario works quite well and more complex models (e.g., linear or quadratic 
microlensing variations) are not required. We performed $\chi^2_{\rm{r,A}}$ and 
$\chi^2_{\rm{r,B}}$ minimizations with a step function--like microlensing for the 10$^4$ 
pairs of simulated curves, yielding delay and magnitude--offset distributions that appear 
in the top panels of Figure~\ref{fig:f5}. The magnitude--offset histograms from the 
$D^2_{4,2}$, $\chi^2_{\rm{r,A}}$, and $\chi^2_{\rm{r,B}}$ estimators are practically 
identical, and thus, indeed, we only include results from $\chi^2_{\rm{r,A}}$ in the top 
right panel of Figure~\ref{fig:f5}.

\begin{deluxetable}{cc}[h!]
\tablecaption{Time delay of SDSS J1442+4055.\label{tab:t5}}
\tablenum{5}
\tablewidth{0pt}
\tablehead{
\colhead{Method} & \colhead{$\Delta t_{\rm{AB}}$} 
}
\startdata
$\chi^2_{\rm{r,A}}$			&			25.2$^{+1.1}_{-1.7}$  \\
$\chi^2_{\rm{r,B}}$ 			&			26.5 $\pm$ 2.3        \\
$D^2_{4,2}$					&			26.3$^{+3.5}_{-1.2}$  \\	  
\enddata
\tablecomments{$\Delta t_{\rm{AB}}$ in days. A is leading, and all measurements are 68\% 
confidence intervals.}
\end{deluxetable}

From the delay distributions in the top left panel of Figure~\ref{fig:f5}, we obtained the 
1$\sigma$ measurements (68\% confidence intervals) in Table \ref{tab:t5}. The 
$\chi^2_{\rm{r,B}}$ and $D^2_{4,2}$ methods provide delay histograms incorporating a 
secondary peak at 29--30 days, which is probably an artefact due to the use of the light 
curve of B as a template for variability ($\chi^2_{\rm{r,B}}$) or related to not 
differentiating between the role that A and B play ($D^2_{4,2}$). Contrarily, the 
$\chi^2_{\rm{r,A}}$ estimator does not provide significant signal at 29--30 days. This 
last technique relies on the use of the light curve of A as a reference template and the 
binned light curve of B, and it is expected to yield the least biased results (errors in A 
are about one half than those in B and noise is reduced when binning original data). In 
addition, selecting the method that produces the smallest uncertainty is a reasonable 
option \citep[e.g.,][]{2013A&A...553A.120T}. Therefore, we adopted a delay interval of 25.0 
$\pm$ 1.5 days (23.5--26.5 days). 

Regarding the magnitude offsets, we derived the 1$\sigma$ intervals: 
$\Delta m_{\rm{AB}}$(2016) = 0.934 $\pm$ 0.002 mag, $\Delta m_{\rm{AB}}$(2017) = 0.956 
$\pm$ 0.002 mag, and $\Delta m_{\rm{AB}}$(2018) = 0.966 $\pm$ 0.003 mag, so we 
unambiguously detected microlensing--induced magnification gradients of 0.022 $\pm$ 0.003 
mag (between 2016 and 2017) and 0.010 $\pm$ 0.004 mag (between 2017 and 2018). From the 
central values in the time delay and magnitude offset intervals, we plotted the combined 
light curve in the $r$ band, i.e., the A brightness record and the magnitude-- and 
time--shifted light curve of B are drawn together (see the bottom panel of 
Figure~\ref{fig:f5}). We remark that exclusively including seasonal changes in the 
$r$--band magnification ratio $\Delta m_{\rm{AB}}$, the shapes of $m_{\rm{A}}(t)$ and 
$m_{\rm{B}}(t + \Delta t_{\rm{AB}}) - \Delta m_{\rm{AB}}$ agree well each other. 

\section{Dust extinction in a high--$z$ galaxy} \label{sec:dust}

\begin{figure}[h!]
\begin{minipage}[h]{0.5\linewidth}
      \centering
      \includegraphics[width=1.0\textwidth]{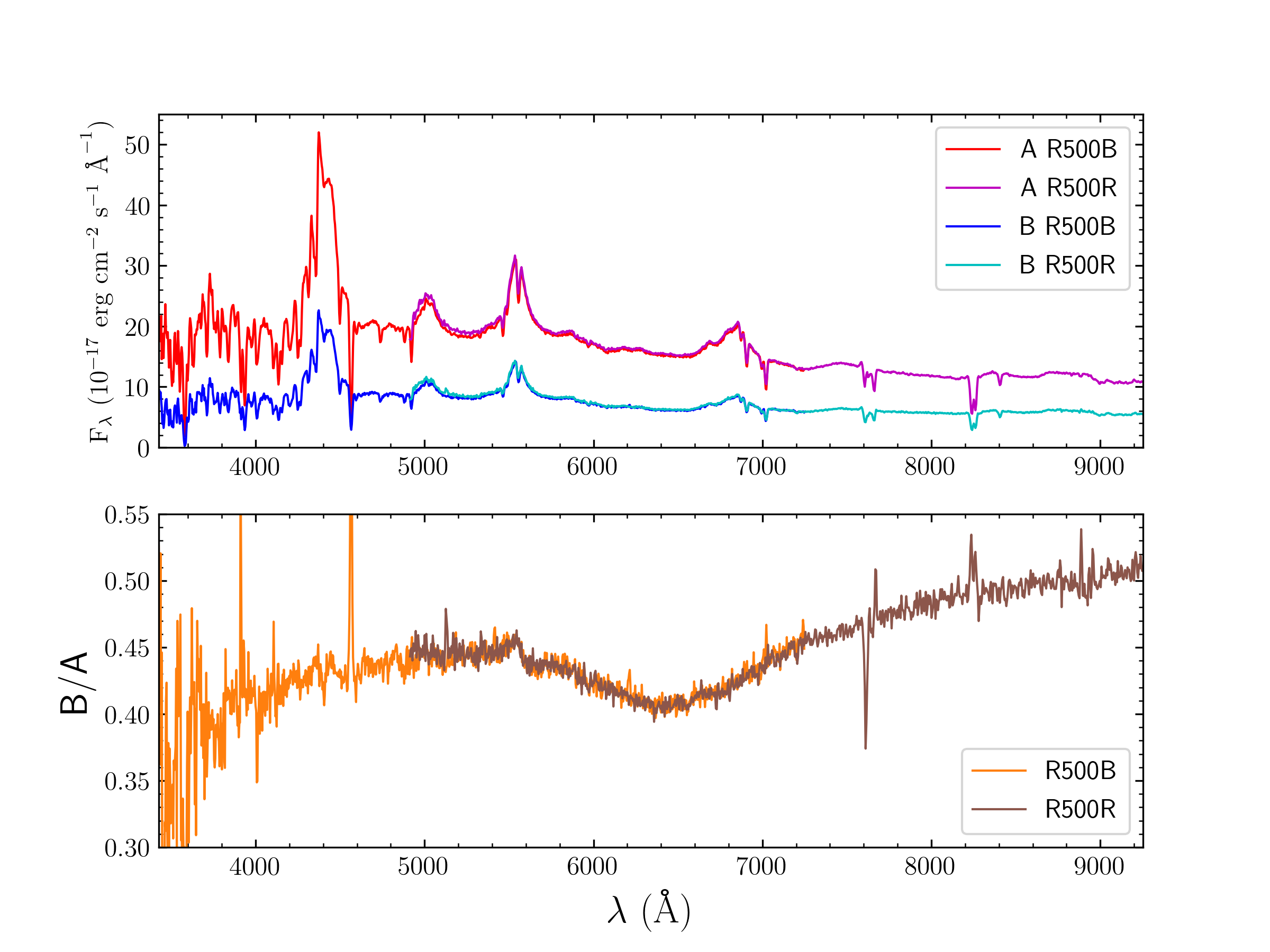}
      \end{minipage}
	\begin{minipage}[h]{0.5\linewidth}
      \centering
      \includegraphics[width=1.0\textwidth]{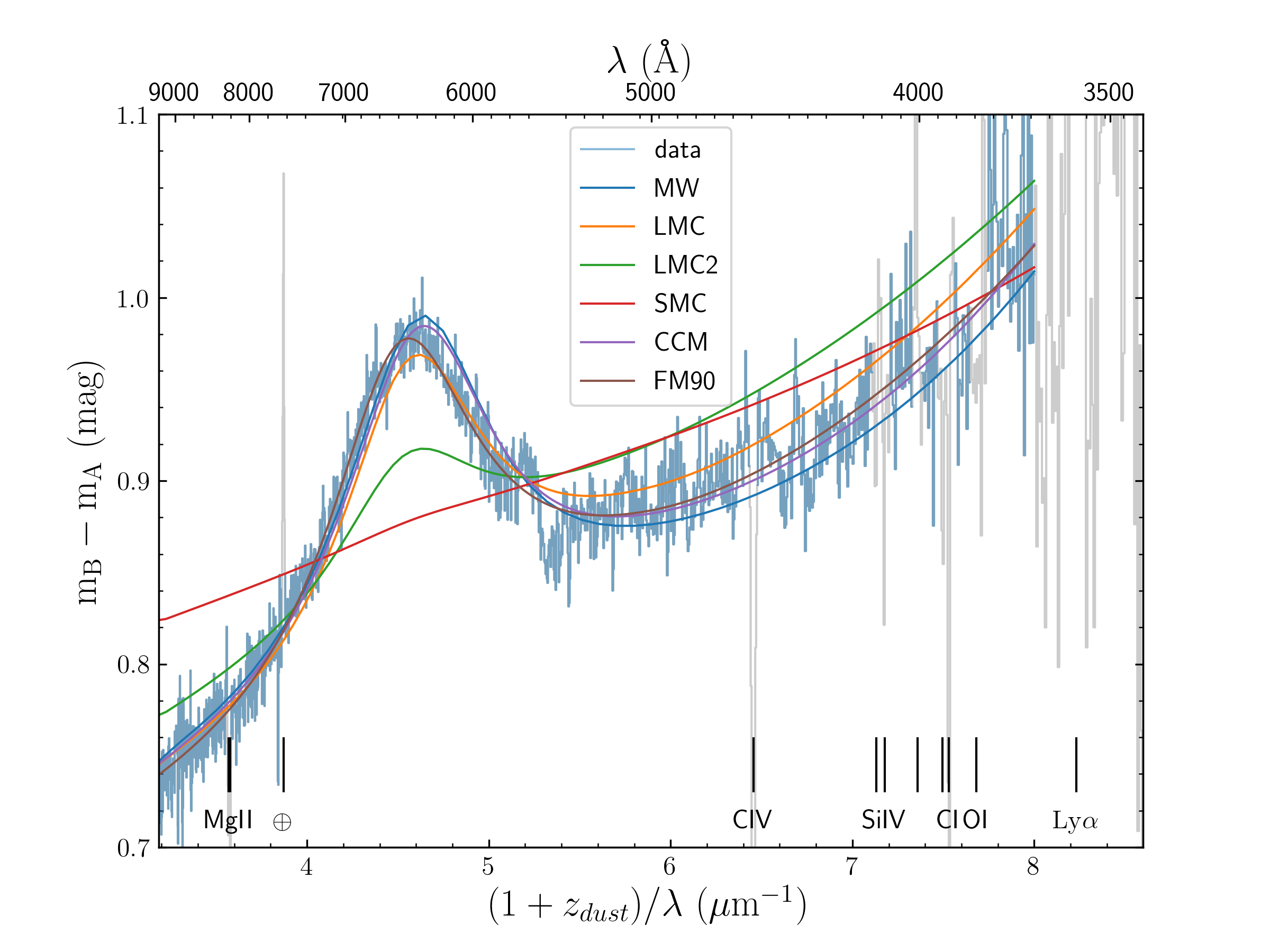}
    	\end{minipage}
\caption{Flux ratios and magnitude differences from the GTC--OSIRIS spectra. Left: Full 
spectra of the quasar images by combining the R500B and R500R data (top), and the 
corresponding flux ratios between 3430 and 9250 \AA\ (bottom). Right: Magnitude differences 
(data) and best fits of different extinction laws. We consider dust at $z_{\rm{dust}}$ = 
1.9465 and six extinction scenarios: average MW (MW), average LMC (LMC), average LMC2 
supershell (LMC2), average SMC bar (SMC), general MW (CCM), and FM90 (see main text for 
details). Very noisy regions (grey line; see the small vertical bars indicating absorption 
features) are not fitted to extinction laws.}
\label{fig:f6} 
\end{figure}

For a double quasar, spectroscopic observations at two epochs separated by approximately 
the time delay between its two images lead to delay--corrected flux ratios at different 
wavelengths, which are valuable tools to study the macro-- and micro--lens magnification 
ratios, as well as the differential dust extinction \citep[e.g.,][]{2006glsw.conf.....M}. 
Our LT $r$--band monitoring of \object{SDSS J1442+4055} yields a relatively short delay of 
about 25 days (A is leading; see Section \ref{sec:delay}), and we have checked that 25 days 
after the GTC--OSIRIS observations, the LT $r$--band flux of B only increased by $\sim$1\%.  
Additionally, the LT $r$--band flux of A at the GTC--OSIRIS observing epoch also increased 
by $\sim$1\% compared to its value 25 days before (see Figure~\ref{fig:f2}). Thus, 
considering calibration uncertainties and $eF_{\lambda}/F_{\lambda}$ values at red 
wavelengths, GTC--OSIRIS single--epoch flux ratios in the red spectral region seem to be 
plausible tracers of those corrected by intrinsic variability. Despite the spectra of SDSS 
J1442+4055AB were taken on a single night, we assumed that these data allow us to build 
flux ratios $B/A$ describing reasonably well the delay--corrected ones. In the left panel 
of Figure~\ref{fig:f6}, we present the single--epoch flux ratios from the GTC--OSIRIS 
spectra, which can be compared with the $A^{\star}/B^{\star}$ values in the top panel of 
Fig. 2 of \citet{2018A&A...619A.142K}. Here, contamination by light from G1 is denoted with 
a star superscript. 

The $B/A$ data are very noisy at the shortest wavelengths, i.e., on the blue edge of the 
R500B grism (see the bottom sub--panel in the left panel of Figure~\ref{fig:f6}), which is 
partially due to the presence of a forest of absorption lines. Additional absorption 
features at longer wavelengths also produce spikes in $B/A$. However, in spectral intervals 
associated with broad line emitting regions, no significant deviations are found with 
respect to adjacent continuum flux ratios. This suggests that chromatic microlensing is 
absent, since compact and extended emitting regions are magnified likewise. Therefore, we 
adopted a constant lens magnification ratio (including both a macro--lens effect caused by 
the entire mass of the gravitational deflectors, and a micro--lens effect produced by stars 
in intervening galaxies), so the chromatic behaviour of $B/A$ is interpreted as due to dust 
extinction. 

Apart from absorption--induced artefacts, the most dramatic feature in the flux ratio 
profile is a broad valley around $\lambda \sim$ 6400 \AA, which is likely related to the 
2175 \AA\ extinction bump seen in some galaxies of the Local Group 
\citep[e.g.,][]{2003ApJ...594..279G}, lensing galaxies at $z \sim$ 1 
\citep[e.g.,][]{2005ApJ...619..749M}, and several metal--rich absorbers at $z \sim$ 2
\citep[e.g.,][]{2017MNRAS.472.2196M}. The existence of this bump is critical to decide 
about the redshift of the intervening dust. Thus, we roughly obtained $z_{\rm{dust}} \sim$ 
1.94, in good agreement with $z_{\rm{IMS}}$. The main lensing galaxy G1 does not seem to 
play a relevant role in extinction, and the high--$z$ IMS would be the main responsible for 
the chromaticity observed in $B/A$. In fact, there is no appreciable absorption at 
$z_{\rm{G1}}$ = 0.284 in the quasar spectra, while both spectra are clearly absorbed at 
$z_{\rm{IMS}}$ = 1.9465. Figure~\ref{fig:f7} does not show any significant distortion of 
the Ly$\alpha$ profiles at $\lambda$ = 3590 \AA, where the Mg\,{\sc ii} 2796 line would 
have been seen if Mg\,{\sc ii} absorption had occurred in G1.

We converted flux ratios into magnitude differences, $m_{\rm{B}} - m_{\rm{A}} = -2.5 \log 
(B/A)$, and then used several extinction laws to fit these differences 
\citep[e.g.,][]{1999ApJ...523..617F,2006ApJS..166..443E}. For $z_{\rm{dust}}$ = 1.9465, we 
are probing the UV extinction in the distant dusty galaxy, i.e., at rest--frame wavelengths 
between 1165 and 3140 \AA. This spectral range practically coincides with the wavelength 
coverage of the low--dispersion mode of the {\it International Ultraviolet Explorer} 
satellite, which provided a wealth of data on the UV extinction of close stars 
\citep[e.g.,][]{1990ApJS...72..163F}. According to concordance cosmology with $H_0$ = 70 km 
s$^{-1}$ Mpc$^{-1}$, $\Omega_m$ = 0.27, and $\Omega_{\Lambda}$ = 0.73
\citep{2009ApJS..180..330K}, the transverse distance between the two light paths at 
$z_{\rm{dust}}$ \citep{1992ApJ...389...39S,2010MNRAS.409..679C} is only $\sim$0.7 kpc 
\citep[see also][]{2018A&A...619A.142K}. We implicitly assumed that dust properties are the 
same along the lines of sight to both images. In addition, we did not fit noisy magnitude 
differences at the shortest wavelengths nor a series of spikes caused by absorption features 
at longer wavelengths (see the right panel of Figure~\ref{fig:f6}). 

First, we considered the average extinction curve of the Milky Way 
\citep[MW;][]{1989ApJ...345..245C}, as well as average curves for other objects in the 
Local Group \citep{2003ApJ...594..279G}: Large Magellanic Cloud (LMC), LMC2 supershell 
(LMC2), and Small Magellanic Cloud bar (SMC). Our data are quite inconsistent with the 
average extinction curves of the LMC2 and SMC (green and red lines in the right panel of 
Figure~\ref{fig:f6}), and show a behaviour halfway between the average curves for the LMC and 
MW (orange and blue lines in the right panel of Figure~\ref{fig:f6}). Second, we obtained a 
significant improvement in the reduced chi--square value when fitting a general Galactic 
extinction law \citep[][hereafter CCM]{1989ApJ...345..245C}. This CCM relationship led to a 
lens magnification ratio of 0.490 $\pm$ 0.005 mag (the constant term in $m_{\rm{B}} - 
m_{\rm{A}}$), a differential visual extinction $\Delta A_{\rm{AB}}(V) = A_{\rm{B}}(V) - 
A_{\rm{A}}(V)$ of 0.133 $\pm$ 0.003 mag, and a total--to--selective extinction ratio $R(V)$ 
of 2.672 $\pm$ 0.048 ($\chi^2_{\rm{r}}$ = 3.77; purple line in the right panel of 
Figure~\ref{fig:f6}).
 
As a final step, in order to accurately describe the observed bump, data were fitted by the 
wavelength--dependent function of \citet[][hereafter FM90]{1990ApJS...72..163F}. FM90 
introduced a Drude (Lorentzian--like) profile for representing a bump with central 
wavenumber $x_0 = 1/\lambda_0$ and width (FWHM) $\gamma$, and $x_0$ = 4.527 $\pm$ 0.004 
$\mu$m$^{-1}$ and $\gamma$ = 0.99 $\pm$ 0.02 $\mu$m$^{-1}$ were obtained from the fit 
($\chi^2_{\rm{r}}$ = 3.30; brown line in the right panel of Figure~\ref{fig:f6}). In the 
right panel of Figure~\ref{fig:f6}, the residuals of the purple and brown lines have 
amplitudes similar to those of the observed noise. Hence, although our best 
$\chi^2_{\rm{r}}$ values for the CCM and FM90 extinction laws are clearly greater than one, 
formal uncertainties in magnitude differences may be underestimated by a factor $\sim$2. 
While the value of $\gamma$ is typical for sight lines towards Galactic stars, the central 
wavelength of the extinction bump ($\lambda_0$ = 2209 $\pm$ 2 \AA) is extraordinarily 
unusual in the MW \citep[e.g.,][]{2007ApJ...663..320F}. However, values of $x_0$ close to 
4.53 $\mu$m$^{-1}$ are consistent with measurements in the LMC 
\citep[e.g.,][]{2003ApJ...594..279G} and in some metal--rich absorbers at $z \sim$ 1--2 
\citep[e.g.,][]{2017MNRAS.472.2196M}.       

\section{High--$z$ gas and its correlation with dust} \label{sec:gas}

\subsection{Complementary observations} \label{subsec:sdssmmt}

To analyse the gas content of the IMS is of interest not only the use of the 
low--resolution GTC--OSIRIS spectra of both quasar images (resolving power of 
$\sim$300--400; see Section \ref{subsec:osiris}), but also other available, not previously 
analysed, medium--resolution spectroscopic data. This higher resolution allows to identify 
finer spectral details, e.g., resolve blended absorption lines. Therefore, in addition to 
the GTC--OSIRIS data, we used the SDSS--BOSS spectrum of the A image with a resolving power 
of $\sim$2000, as well as the data that were obtained at the MMT Observatory with the Blue 
Channel Spectrograph \citep{2018ApJS..236...44F}. The MMT spectra of A and B on 2015 June 
14 cover a wavelength range of 3500--5500 \AA\ at spectral resolution of $\sim$1800, which 
is about 5 times higher than those of the R500B and R500R grisms. For each spectrum, 
we fitted a global continuum and obtained normalized fluxes using the Linetools 
software\footnote{Linetools is a Python package mainly aimed at the identification and 
analysis of absorption lines in quasar spectra. This is publicly available at 
\url{https://github.com/profxj/linetools}}. 

\subsection{Neutral hydrogen} \label{subsec:Hi}

\begin{figure}[h!]
\begin{minipage}[h]{0.5\linewidth}
      \centering
      \includegraphics[width=1.0\textwidth]{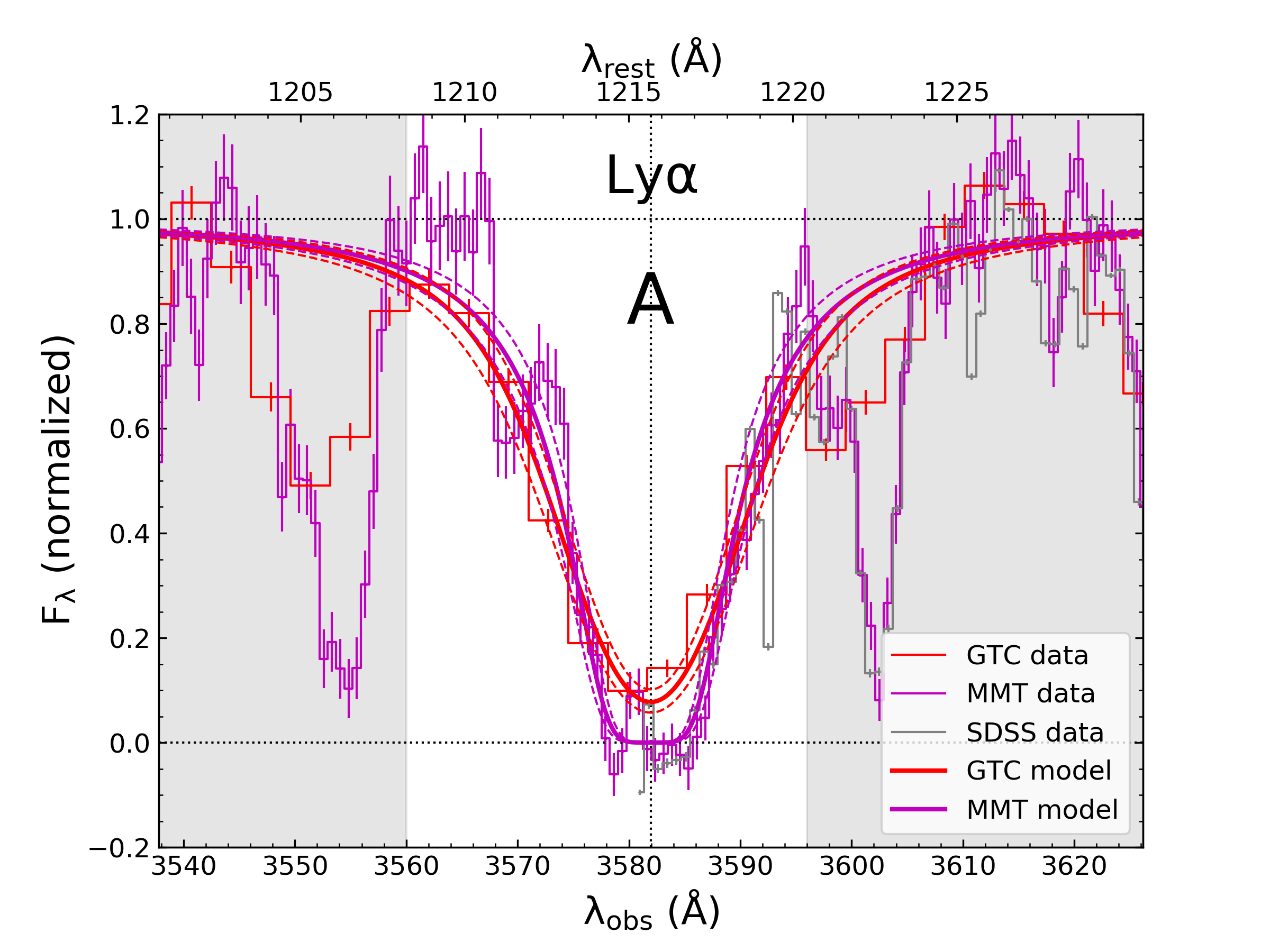}
      \end{minipage}
	\begin{minipage}[h]{0.5\linewidth}
      \centering
      \includegraphics[width=1.0\textwidth]{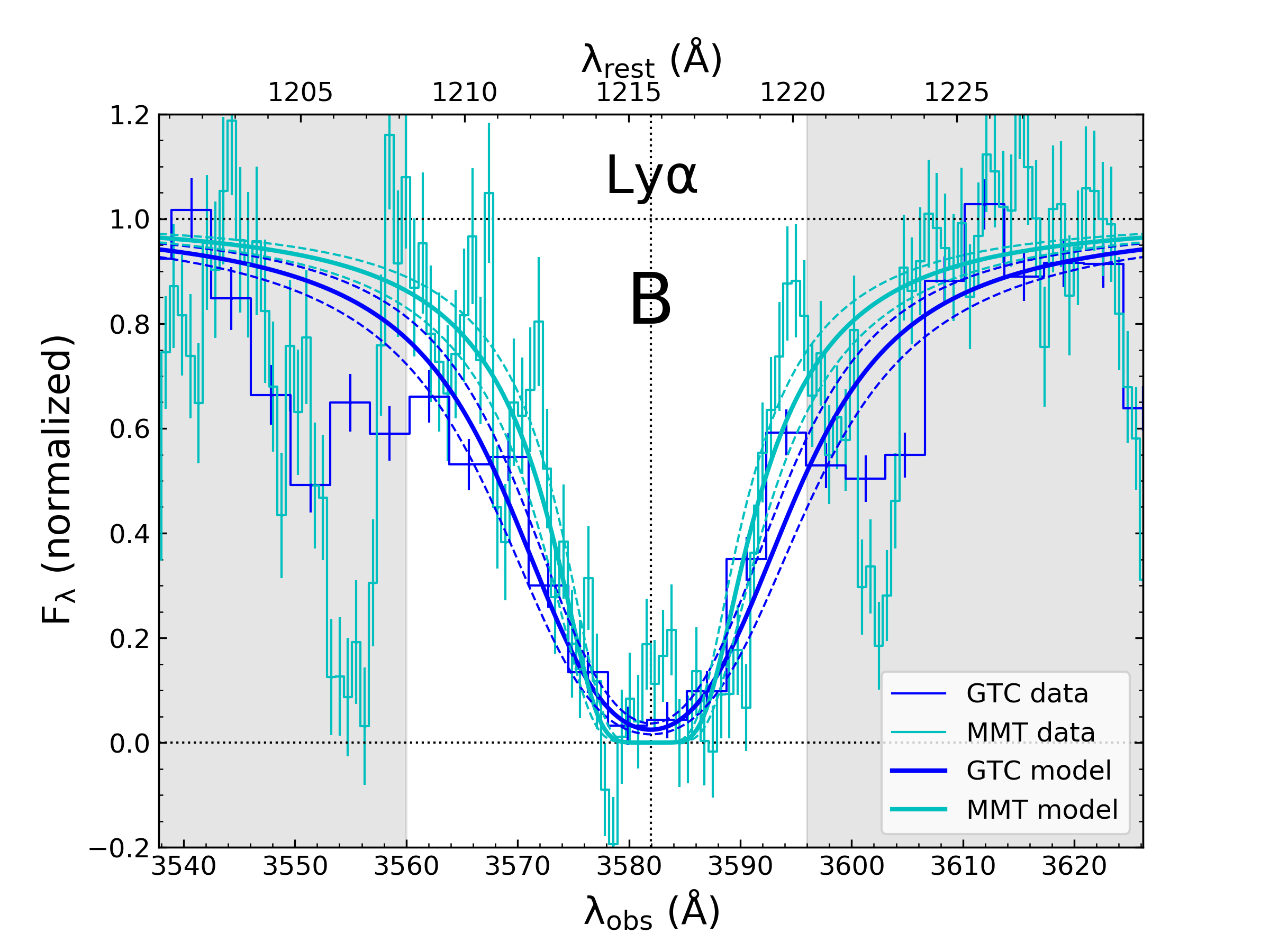}
    	\end{minipage}
\caption{Ly$\alpha$ line profiles of A and B from the GTC--OSIRIS and MMT spectra. The 
thick solid lines are the Voigt functions (convolved with Gaussian instrumental responses) 
that best fit the GTC--OSIRIS and MMT normalized fluxes in the non--shaded wavelength 
interval, and the dashed lines represent their 1$\sigma$ uncertainties. Left: Profiles of 
the image A. The SDSS normalized fluxes are included only for comparison purposes, since 
the SDSS--BOSS spectrum exclusively covers the red wing of the Ly$\alpha$ line. Right: 
Profiles of the image B.}
\label{fig:f7}  
\end{figure}

The GTC--OSIRIS and MMT spectra cover the Ly$\alpha$ absorption at $z_{\rm{gas}}$ = 1.9465,
which is observed around 3582 \AA. Regarding the GTC--OSIRIS spectra, the Ly$\alpha$ line 
profile of the B image is deeper and wider than that of the A image, and this suggests a 
larger H\,{\sc i} column density along the line of sight to B. Using both data sets at 
different spectral resolutions, we fitted line profiles to a Voigt function convolved with 
a Gaussian instrumental profile \citep{2018arXiv180301187K}. These fits were performed with 
the VoigtFit software\footnote{VoigtFit is a Python package for Voigt profile fitting that 
is publicly available at \url{https://github.com/jkrogager/VoigtFit}}. In 
Figure~\ref{fig:f7}, we show the best fits (thick solid lines) along with their 1$\sigma$ 
uncertainties (dashed lines). We note that fits were done by minimising $\chi^2$ in the 
interval 3560 $\leq \lambda_{\rm{obs}} \leq$ 3596 \AA\ (central, non--shaded region in the 
two panels of Figure~\ref{fig:f7}), so that we avoided the Si\,{\sc iii} 1206 line and 
another prominent absorption feature at $\lambda_{\rm{obs}}$ = 3603 \AA. Furthermore, in 
order to estimate 1$\sigma$ confidence intervals, we used 1000 repetitions of each 
Ly$\alpha$ profile. To obtain a repetition of an original Ly$\alpha$ profile, we modified 
the normalized observed fluxes by adding realizations of normal distributions around zero, 
with standard deviations equal to the measured errors. 

The GTC--OSIRIS and MMT data yield the neutral--hydrogen 
column densities in Table \ref{tab:t6}. It is evident that both measures of $\log 
N_{\rm{B}}$(H\,{\sc i}) differ by $\sim$ 0.2, which is an order of magnitude larger than 
formal errors. Thus, we adopted a statistical approach, considering the two values in Table 
\ref{tab:t6} (20.490 and 20.279) and $\log N_{\rm{B}}$(H\,{\sc i}) from the rest--frame 
equivalent width (EW) of the Ly$\alpha$ line in the MMT spectrum of B \citep[20.26; see Eq. 
(9.24) of][]{2011piim.book.....D}. Calculating the average value and its standard deviation, 
and taking into account that the standard deviation of the mean of three values is 50\% 
uncertain, we obtain $\log N_{\rm{B}}$(H\,{\sc i}) = 20.34 $\pm$ 0.11. From the two values 
of $\log N_{\rm{A}}$(H\,{\sc i}) in Table \ref{tab:t6}, we also infer $\log 
N_{\rm{A}}$(H\,{\sc i}) = 20.14 $\pm$ 0.11, where the error of the mean was conservatively 
enlarged to 0.11. The IMS can be classified as a sub--damped/damped Ly$\alpha$ (subDLA/DLA) 
system \citep{1986ApJS...61..249W}, and our H\,{\sc i} column densities agree (although 
having larger uncertainties) with those from high--resolution Keck--HIRES spectra of the 
quasar at $\lambda <$ 6000 \AA\ on 2017 May 20 \citep{2018A&A...619A.142K}. 

\begin{deluxetable}{lcc}[h!]
\tablecaption{Neutral--hydrogen column densities.\label{tab:t6}}
\tablenum{6}
\tablewidth{0pt}
\tablehead{
\colhead{Facility} & 
\colhead{$\log N_{\rm{A}}$(H\,{\sc i})\tablenotemark{a}} & 
\colhead{$\log N_{\rm{B}}$(H\,{\sc i})\tablenotemark{a}} 
}
\startdata
GTC--OSIRIS  & 20.160 $\pm$ 0.021 & 20.490 $\pm$ 0.029 \\
MMT    	 & 20.130 $\pm$ 0.015 & 20.279 $\pm$ 0.025 \\
\enddata
\tablenotetext{a}{Column density $N$ in cm$^{-2}$.}
\end{deluxetable} 

\subsection{Dust--to--gas ratio} \label{subsec:dustgas}

The visual extinction $A(V)$ is proportional to the optical depth at $\lambda_{\rm{rest}}$ 
= 0.55 $\mu$m, which in turn is proportional to the dust grain column density 
$N(\rm{dust})$. Assuming that $N(\rm{dust}) \propto N$(H\,{\sc i}) 
\citep[e.g.,][]{1989ApJ...337....7F,1997ApJ...477..568Z}, we then obtained $A_{\rm{B}}(V)/
A_{\rm{A}}(V)$ = $N_{\rm{B}}$(H\,{\sc i})/$N_{\rm{A}}$(H\,{\sc i}) $\sim$ 1.6 (see Section 
\ref{subsec:Hi}). From this visual extinction ratio and the differential visual extinction 
in Section \ref{sec:dust}, it is possible to estimate the effect of dust along each line of 
sight: $A_{\rm{A}}(V) \sim$ 0.22 mag and $A_{\rm{B}}(V) \sim$ 0.35 mag. We remark the 
similarity between these individual extinctions and the $A(V)$ values found in   
\citet{2018A&A...619A.142K}. The colour excesses of the individual images would be 
$E_{\rm{A}}(B-V) \sim$ 0.08 mag and $E_{\rm{B}}(B-V) \sim$ 0.13 mag, and thus, the bump 
strength (area of the extinction bump) may be estimated at 0.47 and 0.76 mag $\mu$m$^{-1}$ 
for A and B, respectively. These strengths agree well with those of the LMC and metal--rich 
absorbers at $z \sim$ 1--2, whereas are weaker than most measures in the MW \citep[see Fig. 
1 of][]{2017MNRAS.472.2196M}. 

The ratio between $A(V)$ (or $E(B-V)$) and $N$(H\,{\sc i}) is usually called the 
dust--to--gas ratio \citep[e.g.,][and references therein]{2018MNRAS.474.4870M}. For the IMS 
of \object{SDSS J1442+4055}, we derived $A(V)/N$(H\,{\sc i}) $\sim$ 1.6 $\times$ 10$^{-21}$ 
mag cm$^2$, and such a high value is also observed in some absorbers with high metallicity 
(see Section \ref{subsec:metaldust}). This dust--to--gas ratio is a factor of $\sim$ 3 
higher than that of the local interstellar medium \citep{2014ApJ...783...17L} and the MW 
average visual extinction per H for $R(V) \sim$ 2.7 \citep[see Fig. 3 
of][]{2003ARA&A..41..241D}, as well as about 5 times higher than the mean ratio of the LMC 
and Mg\,{\sc ii} absorbers \citep{2003ApJ...594..279G,2009MNRAS.393..808M}. Moreover, the 
mean ratio of high--$z$ DLAs is nearly two orders of magnitude lower than the 
$A(V)/N$(H\,{\sc i}) value for the IMS \citep{2008A&A...478..701V}. Lastly, it is worth 
mentioning that $E(B-V)/N$(H\,{\sc i}) $\sim$ 6 $\times$ 10$^{-22}$ mag cm$^2$, which is 
similar to the corresponding ratio of the distant lensing galaxy of SBS 0909+532 
\citep{2009ApJ...692..677D}. 

\subsection{Metals} \label{subsec:metals}

\begin{figure}
\centering
\includegraphics[width=1.0\textwidth]{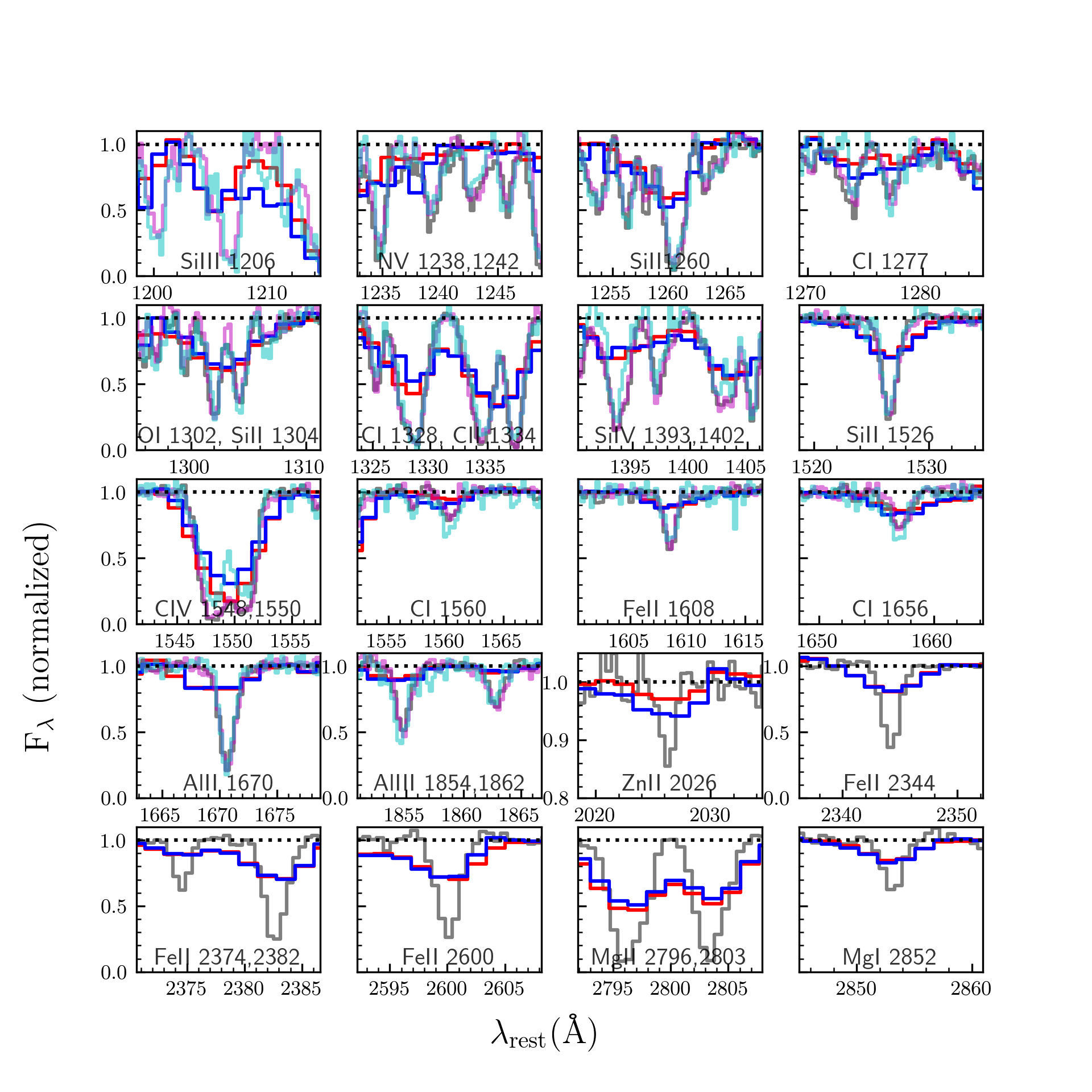}
\caption{Prominent metal absorption lines due to the IMS towards SDSS J1442+4055. Profiles 
of A are marked in grey (SDSS--BOSS data), red (GTC--OSIRIS data), and magenta (MMT data), 
and profiles of B are marked in blue (GTC--OSIRIS data) and cyan (MMT data).} 
\label{fig:f8}
\end{figure}

\begin{deluxetable}{ccccc}
\tablecaption{Rest--frame equivalent widths of main absorption lines produced by the 
IMS.\label{tab:t7}}
\tablenum{7}
\tablewidth{0pt}
\tablehead{
\colhead{Ion} & 
\colhead{$\lambda_{\rm{rest}}$\tablenotemark{a}} & 
\colhead{EW$_{\rm{A}}$\tablenotemark{b}} & 
\colhead{EW$_{\rm{B}}$\tablenotemark{b}} &
\colhead{Data} 
}
\startdata
Si\,{\sc iii} & 1206.50 & 2.183 $\pm$ 0.080 & 1.747 $\pm$ 0.105 & MMT \\
H\,{\sc i}    & 1215.67 & 8.137 $\pm$ 0.215 & 9.899 $\pm$ 0.420 & MMT \\
N\,{\sc v}    & 1238.82 & 0.777 $\pm$ 0.053 & 0.830 $\pm$ 0.086 & MMT \\
N\,{\sc v}    & 1242.80 & 0.768 $\pm$ 0.046 & 0.588 $\pm$ 0.076 & MMT \\
S\,{\sc ii}   & 1253.80 & 0.767 $\pm$ 0.040 & 0.873 $\pm$ 0.065 & MMT \\
Si\,{\sc ii}  & 1260.42 & 1.910 $\pm$ 0.037 & 1.801 $\pm$ 0.058 & MMT \\
C\,{\sc i}    & 1277.24 & 0.723 $\pm$ 0.048 & 0.666 $\pm$ 0.060 & MMT \\
O\,{\sc i}    & 1302.17 & 0.963 $\pm$ 0.040 & 1.090 $\pm$ 0.135 & MMT \\
Si\,{\sc ii}  & 1304.37 & 0.767 $\pm$ 0.035 & 0.815 $\pm$ 0.071 & MMT \\
C\,{\sc i}    & 1328.83 & 2.622 $\pm$ 0.052 & 2.555 $\pm$ 0.089 & MMT \\
C\,{\sc ii}   & 1334.53 & 2.516 $\pm$ 0.048 & 2.236 $\pm$ 0.074 & MMT \\
Si\,{\sc iv}  & 1393.78 & 1.880 $\pm$ 0.043 & 1.269 $\pm$ 0.062 & MMT \\
Si\,{\sc iv}  & 1402.77 & 1.856 $\pm$ 0.040 & 1.455 $\pm$ 0.063 & MMT \\
Si\,{\sc ii}  & 1526.70 & 1.286 $\pm$ 0.034 & 1.099 $\pm$ 0.034 & MMT \\
C\,{\sc iv}   & 1548.20 & 2.627 $\pm$ 0.040 & 1.998 $\pm$ 0.058 & MMT \\ 
C\,{\sc iv}   & 1550.77 & 2.238 $\pm$ 0.039 & 1.761 $\pm$ 0.055 & MMT \\
C\,{\sc i}    & 1560.30 & 0.287 $\pm$ 0.031 & 0.462 $\pm$ 0.038 & MMT \\
Fe\,{\sc ii}  & 1608.45 & 0.565 $\pm$ 0.018 & 0.604 $\pm$ 0.030 & MMT \\
C\,{\sc i}    & 1656.92 & 0.443 $\pm$ 0.037 & 0.690 $\pm$ 0.049 & MMT \\
Al\,{\sc ii}  & 1670.78 & 1.391 $\pm$ 0.041 & 1.123 $\pm$ 0.038 & MMT \\
Al\,{\sc iii} & 1854.71 & 0.937 $\pm$ 0.031 & 0.764 $\pm$ 0.051 & MMT \\
Al\,{\sc iii} & 1862.78 & 0.587 $\pm$ 0.030 & 0.418 $\pm$ 0.044 & MMT \\
Zn\,{\sc ii}  & 2026.13 & 0.170 $\pm$ 0.038 & 0.318 $\pm$ 0.072 & GTC--OSIRIS \\
Fe\,{\sc ii}  & 2344.21 & 1.527 $\pm$ 0.025 & 1.487 $\pm$ 0.043 & GTC--OSIRIS \\
Fe\,{\sc ii}  & 2382.76 & 1.614 $\pm$ 0.032 & 1.612 $\pm$ 0.057 & GTC--OSIRIS \\ 
Fe\,{\sc ii}  & 2600.17 & 1.824 $\pm$ 0.032 & 1.825 $\pm$ 0.062 & GTC--OSIRIS \\
Mg\,{\sc ii}  & 2796.35 & 3.625 $\pm$ 0.045 & 3.563 $\pm$ 0.077 & GTC--OSIRIS \\
Mg\,{\sc ii}  & 2803.53 & 2.417 $\pm$ 0.035 & 2.189 $\pm$ 0.060 & GTC--OSIRIS \\
Mg\,{\sc i}   & 2852.96 & 0.781 $\pm$ 0.024 & 0.927 $\pm$ 0.045 & GTC--OSIRIS \\
\enddata
\tablenotetext{a}{Central rest--frame wavelength in \AA.}
\tablenotetext{b}{Rest--frame equivalent width in \AA.}
\end{deluxetable}

\begin{figure}
\centering
\includegraphics[width=0.8\textwidth]{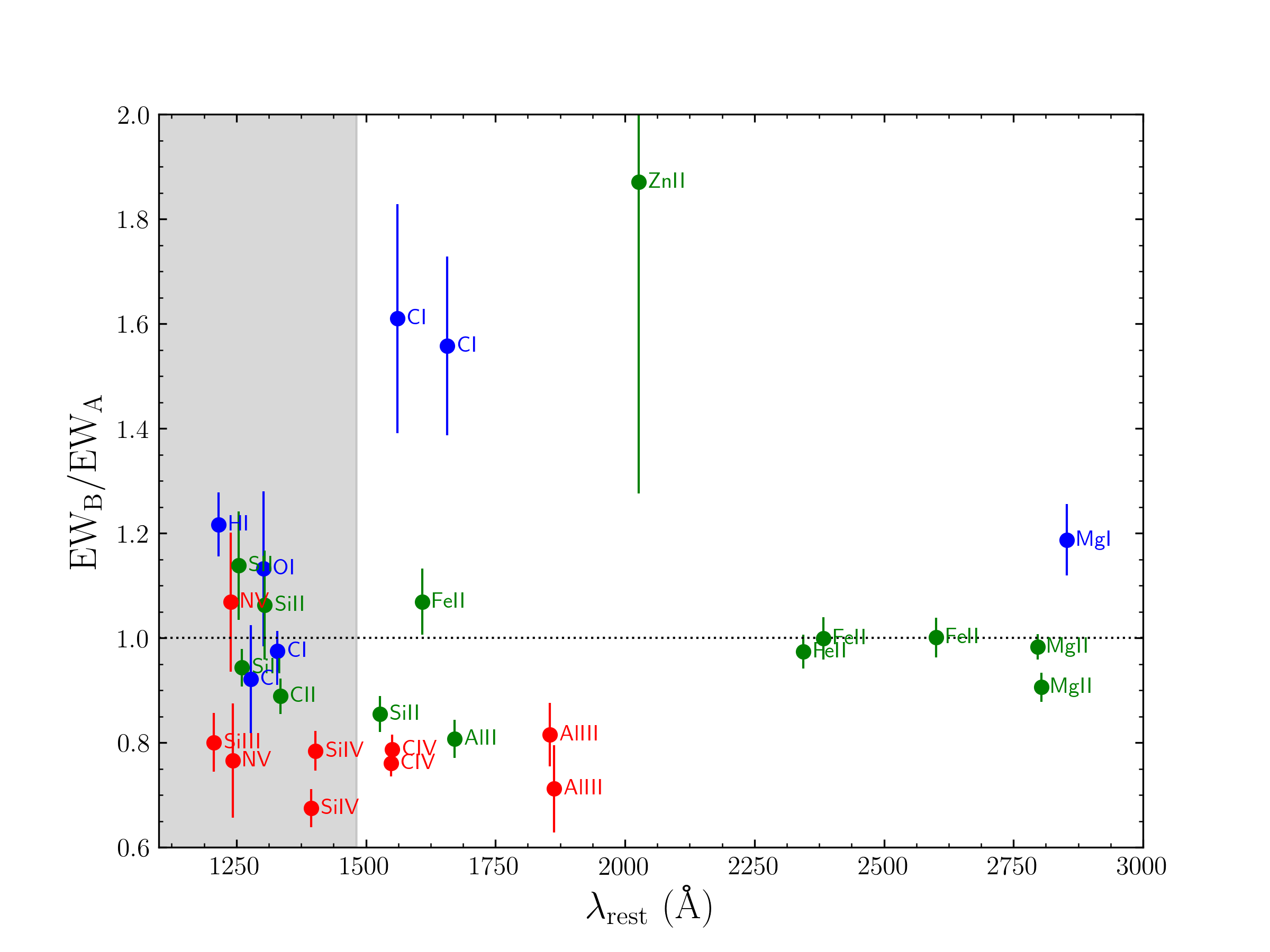}
\caption{EW ratios. We highlight absorption features blueward of the Ly$\alpha$ emission of 
the quasar (grey region), because some of them could be blended with lines belonging to the 
Ly$\alpha$ forest. Blue, green, and red circles denote ratios of neutral atoms, 
low--ionisation metals, and high--ionisation metals, respectively.} 
\label{fig:f9}
\end{figure}

In addition to neutral hydrogen, quasar spectra are affected by metals at $z_{\rm{gas}}$ = 
1.9465. Prominent metal lines are shown in Figure~\ref{fig:f8}, which incorporates 
data from the SDSS--BOSS (grey; only for A), the GTC--OSIRIS (red for A and blue for B), 
and the MMT Observatory (magenta for A and cyan for B). The GTC--OSIRIS line profiles were 
then used to determine rest--frame EWs of absorption lines at $\lambda_{\rm{obs}} \geq$ 
5500 \AA, whereas EWs of lines observed at shorter wavelengths were derived from the MMT 
profiles (see Section \ref{subsec:sdssmmt}). The EWs of main absorption features are listed 
in Table \ref{tab:t7}. The neutral carbon detection is particularly interesting, since the 
appearance of a 2175 \AA\ extinction bump at $z \sim$ 2 has been linked to the presence of 
C\,{\sc i} \citep[e.g.,][]{2009ApJ...697.1725E,2015A&A...580A...8L,2018MNRAS.474.4870M}. We 
observe strong absorption (EW $>$ 1 \AA) of C\,{\sc i} 1328, and weaker lines of C\,{\sc i} 
1277, C\,{\sc i} 1560, and C\,{\sc i} 1656. We also note the presence of Zn\,{\sc ii} 2026. 
Despite the Zn\,{\sc ii} 2026 and Mg\,{\sc i} 2026 lines are blended with each other, and 
we initially measured EW(Zn\,{\sc ii} 2026 + Mg\,{\sc i} 2026) for A and B, EW(Zn\,{\sc ii} 
2026) values are presented in Table \ref{tab:t7}. We used the correction EW(Zn\,{\sc ii} 
2026) = EW(Zn\,{\sc ii} 2026 + Mg\,{\sc i} 2026) $-$ [$f$(Mg\,{\sc i} 2026)/$f$(Mg\,{\sc i} 
2852)] $\times$ EW(Mg\,{\sc i} 2852), where $f$(Mg\,{\sc i} 2026) and $f$(Mg\,{\sc i} 2852) 
are oscillator strengths.

As a first approximation, EW$_{\rm{B}}$/EW$_{\rm{A}}$ ratios were used to estimate relative 
amounts of neutral hydrogen and metals along the two sight lines (see Figure~\ref{fig:f9}). 
These quantities approximately represent column density ratios $N_{\rm{B}}$/$N_{\rm{A}}$ at
low optical depths, but they describe $(N_{\rm{B}}$/$N_{\rm{A}})^{1/2}$ values at very high 
optical depths \citep[e.g.,][]{2011piim.book.....D}. Thus, in Figure~\ref{fig:f9}, we have 
to remark that EW$_{\rm{B}}$(H\,{\sc i})/EW$_{\rm{A}}$(H\,{\sc i}) $\sim$ 
$[N_{\rm{B}}$(H\,{\sc i})/$N_{\rm{A}}$(H\,{\sc i})$]^{1/2}$. Section \ref{subsec:Hi} 
provides data for a reliable estimation of $N_{\rm{B}}$(H\,{\sc i})/$N_{\rm{A}}$(H\,{\sc 
i}). In Figure~\ref{fig:f9}, we differentiate between neutral atoms (blue circles), 
low--ionisation metals (green circles), and high--ionisation metals (red circles). 

The C\,{\sc i} 1560 and C\,{\sc i} 1656 absorption lines (see Table \ref{tab:t7}) are 
formed at relatively low optical depths, so their EW ratios roughly correspond to column 
density ratios. We thus obtain that the neutral--carbon column density for B (the most 
reddened image) is appreciably higher than for A. From the Keck--HIRES high--resolution 
spectra, \citet{2018A&A...619A.142K} also found a $N_{\rm{B}}$(C\,{\sc 
i})/$N_{\rm{A}}$(C\,{\sc i}) ratio above 3. Even though both images seem to be affected by 
similar amounts of Fe\,{\sc ii} and Mg\,{\sc ii}, the Zn\,{\sc ii} absorption is stronger 
along the line of sight to B. Non-refractory (volatile) elements, e.g., Zn, condensate onto 
dust grains much more difficultly than refractory elements, e.g., Fe and Mg. Hence, in the 
B image, we observe a relative excess of Zn in the gas phase. However, there are no 
gas--phase excesses of Fe and Mg, which are more easily trapped in dust grains. It is also 
clear that the A image is more affected by high--ionisation metals. 

\subsection{Metallicity and dust depletion} \label{subsec:metaldust}

The two absorption lines of Fe\,{\sc ii} 1608 and Zn\,{\sc ii} 2026 are not saturated and 
lie outside the Ly$\alpha$ forest. Thus, we initially used Eq. (9.15) of 
\citet{2011piim.book.....D} to estimate column densities from their EWs. However, if column 
densities are not actually proportional to EWs (optically--thin regime condition is not 
met), $N$ values are underestimated, being underestimate greater when EW is larger. While 
the equivalent widths of the Zn\,{\sc ii} 2026 line do not exceed 0.3 \AA, the EWs of the 
Fe\,{\sc ii} 1608 line reach 0.6 \AA, and we carried out a more detailed analysis of this 
stronger absorption. For the A image, the optically--thin regime approach led to 
$\log N_{\rm{A}}$(Fe\,{\sc ii}) = 14.63 $\pm$ 0.01. Additionally, the SDSS--BOSS spectrum 
of A allowed us to construct the curve of growth for Fe\,{\sc ii}, as well as to fit the 
two relevant parameters: $\log N_{\rm{A}}$(Fe\,{\sc ii}) = 14.71 $\pm$ 0.04 and $b$ = 69 
$\pm$ 3 km s$^{-1}$. We adopted this last interval for $\log N_{\rm{A}}$(Fe\,{\sc ii}), and 
considered a bias of $-$0.08 to correct our initial estimate of $\log N_{\rm{B}}$(Fe\,{\sc 
ii}) through the optically--thin regime approach (its error was also set to 0.04; see Table 
\ref{tab:t8}).  

In Table \ref{tab:t9}, we also present the metal abundances [Fe/H] and [Zn/H], where [X/H] 
= $\log$[$N$(X)/$N$(H)] $-$ $\log$(X/H)$_{\sun}$. We assumed that $N$(Fe) = $N$(Fe\,{\sc 
ii}), $N$(Zn) = $N$(Zn\,{\sc ii}), and $N$(H) = $N$(H\,{\sc i}), and taken solar abundances 
$\log$(X/H)$_{\sun}$ from \citet{2009ARA&A..47..481A}. Using the average value of [Zn/H] as 
a metallicity estimator (Zn basically remains in the gas phase), the IMS has a super--solar 
metallicity [Zn/H]$_{\rm{IMS}}$ = $+$0.27. This means (Zn/H)$_{\rm{IMS}}$ is about 2 times 
(Zn/H)$_{\sun}$. However, we should bear in mind that \citet{2018A&A...619A.142K} detected 
H$_2$ along both lines of sight, so that using the total hydrogen column density instead of
$N$(H\,{\sc i}), $\log N$(H) must be increased by about 1\%. As a result, 
[Zn/H$_{\rm{T}}$]$_{\rm{IMS}} \sim$ +0.07, which confirms the metallicity from sulphur and 
total hydrogen [S/H$_{\rm{T}}$]$_{\rm{IMS}} \sim$ 0, based on high--resolution Keck--HIRES 
spectra. In addition, the [Fe/H] values are significantly less than zero, indicating 
depletion of refractory elements onto dust grains. The abundance ratio of iron to zinc, 
[Fe/Zn], is commonly used as a dust depletion estimator. It measures the depletion of Fe 
from its gas phase to the dust phase. The column densities in Table \ref{tab:t8} and solar 
metal abundances in Table 1 of \citet{2009ARA&A..47..481A} yielded [Fe/Zn]$_{\rm{A}}$ = 
$-$1.14 $\pm$ 0.11 and [Fe/Zn]$_{\rm{B}}$ = $-$1.38 $\pm$ 0.11, where 
$\log$(Fe/Zn)$_{\sun}$ was averaged over its photosphere and meteorite values.     

\begin{deluxetable}{ccc}
\tablecaption{Fe\,{\sc ii} and Zn\,{\sc ii} column densities.\label{tab:t8}}
\tablenum{8}
\tablewidth{0pt}
\tablehead{
\colhead{Ion} & 
\colhead{$\log N_{\rm{A}}$\tablenotemark{a}} & 
\colhead{$\log N_{\rm{B}}$\tablenotemark{a}} 
}
\startdata
Fe\,{\sc ii}  & 14.71 $\pm$ 0.04 & 14.74 $\pm$ 0.04 \\
Zn\,{\sc ii}  & 12.97 $\pm$ 0.10 & 13.24 $\pm$ 0.10 \\ 
\enddata
\tablenotetext{a}{Column density $N$ in cm$^{-2}$.}
\end{deluxetable} 

\begin{deluxetable*}{lcccc}
\tablenum{9}
\tablecaption{Fe and Zn abundances relative to their solar values.\label{tab:t9}}
\tablewidth{0pt}
\tablehead{
\colhead{} & \multicolumn2c{[Fe/H]} & \multicolumn2c{[Zn/H]} \\
\colhead{} & \colhead{A} & \colhead{B} & \colhead{A} & \colhead{B}
}
\startdata
Photosphere & $-$0.93 $\pm$ 0.12 & $-$1.10 $\pm$ 0.12 & $+$0.27 $\pm$ 0.16 & $+$0.34 $\pm$ 0.16 \\  
Meteorites  & $-$0.88 $\pm$ 0.12 & $-$1.05 $\pm$ 0.12 & $+$0.20 $\pm$ 0.15 & $+$0.27 $\pm$ 0.15 \\  
\enddata
\tablecomments{We consider solar photosphere and meteorite metal abundances in Table 1 of 
Asplund et al. (2009).}
\end{deluxetable*}

The IMS of \object{SDSS J1442+4055} belongs to the family of 2175 \AA\ dust absorbers 
(2DAs) that was studied by \citet{2018MNRAS.474.4870M}, who also assumed $N$(H) = 
$N$(H\,{\sc i}). The 2DAs contain C\,{\sc i} absorbing gas with $\log N$(C\,{\sc i}) $>$ 
14.0 (we obtain $\log N$(C\,{\sc i}) $>$ 14.0 from the EWs of the C\,{\sc i} 1656 line in 
Table \ref{tab:t7}), and the subDLAs/DLAs with $\log N$(H\,{\sc i}) $\sim$ 20.0--20.5 
(subset of the 2DA population) have super--solar metallicities (see Table \ref{tab:t9}). 
These 2DAs also show a strong correlation between dust--to--gas ratio and metallicity. For 
the dust--to--gas ratio of the IMS (see Section \ref{subsec:dustgas}), Eq. (7) of 
\citet{2018MNRAS.474.4870M} predicts a high metallicity [Zn/H] $\sim +$0.3, in good 
agreement with the values in Table \ref{tab:t9}. In addition, if we focus on the 2DAs with
$\log N$(H\,{\sc i}) $\sim$ 20.0--20.5, their high depletion levels agree well with our 
measures of [Fe/Zn] (see above). The values of [Fe/Zn]$_{\rm{A}}$ and [Fe/Zn]$_{\rm{B}}$ 
can be used to estimate the stellar mass of the IMS. From the mass--metallicity--redshift 
relation of \citet{2013MNRAS.430.2680M}, and the observed [Zn/H]--[Fe/Zn] relationship in 
2DAs, we derived $M_{\rm{stars}} \sim 2 \times 10^{10}$ M$_{\sun}$. Therefore, the IMS host 
galaxy appears to be a metal--rich and relatively massive object, containing large amounts 
of dust and neutral gas. In this scenario, star formation is likely to occur.

\section{Conclusions} \label{sec:final}

This paper mainly reports on optical follow--up observations of the gravitationally lensed 
quasar \object{SDSS J1442+4055}, using the GTC and the LT. The main lensing galaxy G1 is 
only 1\farcs38 from the brightest quasar image A, and its GTC spectra clearly show the 
Ca\,{\sc ii} $HK$, G band, H$\beta$, and Mg\,{\sc i} $b$ absorption features at 
$z_{\rm{G1}}$ = 0.284 $\pm$ 0.001. The new spectra of G1 with unprecedented quality might 
be used (together with IR spectroscopy) to fit stellar population models in the non--local 
early--type galaxy \citep[e.g.,][]{bruzual03}. This should lead to realistic microlens mass 
functions to generate microlensing magnification maps. The secondary galaxies G2 and G3 are 
located 5\farcs2 and 33\farcs9 from A, and our LT spectra of these two objects yield 
redshifts $z_{\rm{G2}}$ = $z_{\rm{G3}}$ = 0.22 $\pm$ 0.01. Thus, G2 is not physically 
associated with G1, but it is at the same distance as G3.

The LT $r$--band light curves of the two quasar images A and B over 2.7 years of monitoring
display significant variations, which are used to measure a time delay of 25.0 $\pm$ 1.5 
days (1$\sigma$ confidence interval; A is leading). Despite this delay is robustly measured 
to 6\% precision, before using it to estimate cosmological parameters, one must consider a 
possible microlensing--induced contribution \citep{2018MNRAS.473...80T}. To properly 
account for a putative microlensing bias in the time delay estimation, it is required to 
perform numerical simulations. However, there are reasons to think this bias is well below 
the delay uncertainty of 1.5 days. First, we detect microlensing magnification gradients 
$<$ 10$^{-4}$ mag day$^{-1}$ in the $r$ band. Second, the flux ratios $B/A$ from the GTC 
spectra of both quasar images do not show evidence of microlensing inhomogeneous 
magnification, since sources with different shapes/sizes are magnified equally. 

Current observational constraints also allow us to explore simple mass models for 
\object{SDSS J1442+4055}, and thus, compare the predicted delays with the measured one. We
may consider the astrometry in Table 1 of \citet{2016MNRAS.456.1948S} and the lens 
magnification ratio we derive in Section \ref{sec:dust}, i.e., a macrolens flux ratio $B/A$ 
= 0.64 $\pm$ 0.064, where the uncertainty is increased to 10\% to take an unknown microlens 
effect into account. If we fit a singular isothermal ellipsoid (SIE) mass model to these 
observations ($\chi^2 \sim$ 0), the LENSMODEL software \citep{2001astro.ph..2340K,
2010GReGr..42.2151K} produces an Einstein radius, ellipticity (position angle), and time 
delay of 1\farcs073, 0.034 ($-$28\fdg0), and 26 days \citep[adopting the concordance 
cosmology we use in Section \ref{sec:dust};][]{2009ApJS..180..330K}. As the ellipticity of 
the SIE model is quite small, we could also probe a singular isothermal sphere (SIS), where 
the position of the SIS is allowed to vary during the fitting procedure. Through the 
LENSMODEL package we find a solution with $\chi^2_{\rm{r}}$ = $\chi^2$/dof = 2.45/2. The 
lensing mass parameters are 1\farcs078 (Einstein radius) and ($x_{\rm{lens}}$, 
$y_{\rm{lens}}$) = (1\farcs339, $-$0\farcs323), with the lens centre being slightly offset 
($\sim$ 0\farcs02) from the Sergeyev et al.'s position of G1. This SIS model leads to a 
time delay of 25.3 days, which is very close to our central delay value (see above).

The GTC quasar spectra indicate the presence of an intervening metal system at $z \sim$ 2,
and we measure $z_{\rm{IMS}}$ = 1.9465 from strong metal absorption lines in the SDSS--BOSS 
spectrum of A, which has higher resolution than those of the GTC \citep[see 
also][]{2016MNRAS.456.1948S,2018A&A...619A.142K}. Leaving aside absorption features, the 
high SNR spectroscopy with the GTC offers a unique opportunity to analyse the flux ratios 
$B/A$ over the wide wavelength interval between 3430 to 9250 \AA. A prominent extinction 
bump is detected at a redshift similar to that of the distant IMS, so this high--$z$ object
contains dust grains and gas--phase metals. Assuming dust properties are similar along both
sight lines (A and B), we fit extinction curves to the magnitude differences from the 
measured flux ratios. At $z_{\rm{IMS}}$ = 1.9465, the transverse distance beteen A and B is 
less than 1 kpc. In addition, \citet{2008A&A...485..403O} reported that when several images 
of the same quasar are affected by dust extinction, the preferred values of $R(V)$ are 
similar. A general Galactic extinction curve \citep{1989ApJ...345..245C} yields an 
acceptable fit with $R(V)$ = 2.672 $\pm$ 0.048 and a differential visual extinction 
$A_{\rm{B}}(V) - A_{\rm{A}}(V)$ = 0.133 $\pm$ 0.003 mag. Moreover, using the extinction law
of \citet{1990ApJS...72..163F}, we obtain $x_0$ = 4.527 $\pm$ 0.004 $\mu$m$^{-1}$ and 
$\gamma$ = 0.99 $\pm$ 0.02 $\mu$m$^{-1}$ for the central wavenumber and width (FWHM) of the
extinction bump. The value of $x_0$ is very unusual in the Milky Way, but it agrees with 
values in the LMC and metal--rich absorbers at $z \sim$ 1--2 
\citep[e.g.,][]{2003ApJ...594..279G,2007ApJ...663..320F,2017MNRAS.472.2196M}.  

To accurately study the gas content of the high--$z$ IMS and the dust--gas correlation, we 
use the GTC spectra of A and B, as well as higher resolution data from the SDSS--BOSS 
spectroscopy of A and the MMT observations of both quasar images 
\citep{2018ApJS..236...44F}. Assuming that the dust grain column density is proportional to 
the H\,{\sc i} column density, the visual extinction ratio $A_{\rm{B}}(V)/A_{\rm{A}}(V)$ = 
$N_{\rm{B}}$(H\,{\sc i})/$N_{\rm{A}}$(H\,{\sc i}) $\sim$ 1.6 enables us to know how dust 
affects each individual image. For example, we estimate bump strengths of 0.47 (A) and 0.76 
(B) mag $\mu$m$^{-1}$. These are consistent with bump areas in the LMC and metal--rich 
absorbers at $z \sim$ 1--2 \citep[e.g.,][]{2017MNRAS.472.2196M}. The IMS  at $z \sim$ 2 
belongs to the family of dusty absorbers discussed by \citet{2018MNRAS.474.4870M}, since it 
is a metal--strong sub--damped/damped Ly$\alpha$ system with $\log N$(H\,{\sc i}) $\sim$ 
20.0--20.5, contains C\,{\sc i} gas with $\log N$(C\,{\sc i}) $>$ 14.0, and has high values 
of the dust--to--gas ratio $A(V)/N$(H\,{\sc i}) ($\sim$ 1.6 $\times$ 10$^{-21}$ mag 
cm$^2$), the gas--phase metallicity indicator [Zn/H] ($\sim$ +0.3; H $\equiv$ H\,{\sc i}), 
and the dust depletion level ($-$1.5 $<$ [Fe/Zn] $<$ $-$1). Our results in Table 
\ref{tab:t7} and Figure~\ref{fig:f9} can also be used to check the variation in metal--line 
equivalent width over a transverse physical scale of $\sim$ 0.7 kpc 
\citep[e.g.,][]{2017ApJ...851...88K,2018ApJ...859..146R}. Finally, we note that this work 
and a spectroscopic study of \object{SDSS J1442+4055} by \citet{2018A&A...619A.142K} have 
been conducted concurrently but independently. \citet{2018A&A...619A.142K} have used Keck 
spectra and data analysis methods different from ours to obtain results similar to those we 
present here.

\acknowledgments

We thank the anonymous referee for helpful comments that contributed to improving the final version 
of the paper.
The Liverpool Telescope is operated on the island of La Palma by Liverpool John Moores University 
in the Spanish Observatorio del Roque de los Muchachos of the Instituto de Astrofisica de Canarias 
with financial support from the UK Science and Technology Facilities Council. This article is also 
based on observations made with the Gran Telescopio Canarias, installed at the Spanish Observatorio 
del Roque de los Muchachos of the Instituto de Astrof\'{\i}sica de Canarias, in the island of La 
Palma. We thank the staff of both telescopes for a kind interaction before, during and after the 
observations. We also used data taken from the Sloan Digital Sky Survey (SDSS) database. SDSS is 
managed by the Astrophysical Research Consortium for the Participating Institutions of the SDSS 
Collaboration. The SDSS web site is www.sdss.org. Funding for the SDSS has been provided by the 
Alfred P. Sloan Foundation, the Participating Institutions, and national agencies in the U.S. and 
other countries. SDSS acknowledges support and resources from the Center for High-Performance 
Computing at the University of Utah. We are grateful to the SDSS collaboration for doing that 
public database. This research has been conducted in the framework of the Gravitational LENses and 
DArk MAtter (GLENDAMA) project, which was/is supported by the Spanish Department of Research, 
Development and Innovation grant AYA2013-47744-C3-2-P, the MINECO/AEI/FEDER-UE grant 
AYA2017-89815-P, the complementary action "Lentes Gravitatorias y Materia Oscura" financed by the 
SOciedad para el DEsarrollo Regional de CANtabria (SODERCAN S.A.) and the Operational Programme of 
FEDER-UE, and the University of Cantabria.

%

\vspace{5mm}
\facilities{Liverpool:2m(IO:O and SPRAT), GTC(OSIRIS)}


\software{IRAF (\url{http://iraf.noao.edu/}),
		PyRAF (\url{http://www.stsci.edu/institute/software_hardware/pyraf}),
		SciPy (\url{https://www.scipy.org/}),
 		Astropy \citep{2013A&A...558A..33A,2018AJ....156..123A},
		Linetools \citep{prochaska17}, 
		IMFITFITS \citep{1998AJ....115.1377M},
 		VoigtFit \citep{2018arXiv180301187K},  
         	LENSMODEL \citep{2001astro.ph..2340K}
          }


\begin{thebibliography}{}

\bibitem[Ahn et al.(2014)]{2014ApJS..211...17A}
Ahn C.~P., Alexandroff, R., Allende Prieto, C., et al.\ 2014, \apjs, 211, 17 
\bibitem[Anguita et al.(2018)]{2018MNRAS.480.5017A} 
Anguita, T., Schechter, P.~L., Kuropatkin, N., et al.\ 2018, \mnras, 480, 5017 
\bibitem[Asplund et al.(2009)]{2009ARA&A..47..481A} 
Asplund, M., Grevesse, N., Sauval, A.~J., \& Scott, P.\ 2009, \araa, 47, 481 
\bibitem[Astropy Collaboration(2013)]{2013A&A...558A..33A} 
Astropy Collaboration\ 2013, \aap, 558, A33 
\bibitem[Astropy Collaboration(2018)]{2018AJ....156..123A} 
Astropy Collaboration\ 2018, \aj, 156, A123 
\bibitem[Bonvin et al.(2017)]{2017MNRAS.465.4914B} 
Bonvin, V., Courbin, F., Suyu, S.~H., et al.\ 2017, \mnras, 465, 4914 
\bibitem[Bruzual(2003)]{bruzual03} 
Bruzual A., G.\ 2003, XI Canary Islands Winter School of Astrophysics, Galaxies at High Redshift, 
ed. I. P\'erez-Fournon, M. Balcells, F. Moreno-Insertis \& F. S\'anchez (Cambridge, UK: Cambridge 
University Press), 185
\bibitem[Cardelli et al. (1989)]{1989ApJ...345..245C} 
Cardelli, J.~A., Clayton, G.~C., \& Mathis, J.~S.\ 1989, \apj, 345, 245
\bibitem[Cooke et al.(2010)]{2010MNRAS.409..679C} 
Cooke, R., Pettini, M., Steidel, C.~C., et al.\ 2010, \mnras, 409, 679 
\bibitem[Dai \& Kochanek (2009)]{2009ApJ...692..677D} 
Dai, X., \& Kochanek, C.~S.\ 2009, \apj, 692, 677
\bibitem[Dawson et al.(2013)]{2013AJ....145...10D} 
Dawson, K.~S., Schlegel, D.~J., Ahn, C.~P., et al.\ 2013, \aj, 145, A10
\bibitem[Draine(2003)]{2003ARA&A..41..241D} 
Draine, B.~T.\ 2003, \araa, 41, 241 
\bibitem[Draine(2011)]{2011piim.book.....D} 
Draine, B.~T. 2011, Physics of the Interstellar and Intergalactic Medium
(Princeton, Princeton University Press)
\bibitem[El{\'{\i}}asd{\'o}ttir et al.(2006)]{2006ApJS..166..443E} 
El{\'{\i}}asd{\'o}ttir, {\'A}., Hjorth, J., Toft, S., Burud, I., \& Paraficz, D.\ 2006, \apjs, 166, 443
\bibitem[El{\'{\i}}asd{\'o}ttir et al.(2009)]{2009ApJ...697.1725E} 
El{\'{\i}}asd{\'o}ttir, {\'A}., Fynbo, J.~P.~U., Hjorth, J., et al.\ 2009, \apj, 697, 1725 
\bibitem[Ellison et al.(2010)]{2010MNRAS.406.1435E} 
Ellison, S.~L., Prochaska, J.~X., Hennawi, J., et al.\ 2010, \mnras, 406, 1435 
\bibitem[Falco et al. (1999)]{1999ApJ...523..617F} 
Falco, E.~E., Impey, C.~D., Kochanek, C.~S., et al.\ 1999, \apj, 523, 617
\bibitem[Fall \& Pei(1989)]{1989ApJ...337....7F} 
Fall, S.~M., \& Pei, Y.~C.\ 1989, \apj, 337, 7 
\bibitem[Findlay et al.(2018)]{2018ApJS..236...44F} 
Findlay, J.~R., Prochaska, J.~X., Hennawi, J.~E., et al.\ 2018, \apjs, 236, 44
\bibitem[Fitzpatrick \& Massa(1990)]{1990ApJS...72..163F} 
Fitzpatrick, E.~L., \& Massa, D.\ 1990, \apjs, 72, 163 
\bibitem[Fitzpatrick \& Massa(2007)]{2007ApJ...663..320F} 
Fitzpatrick, E.~L., \& Massa, D.\ 2007, \apj, 663, 320 
\bibitem[Gil-Merino et al.(2018)]{2018A&A...616A.118G}
Gil-Merino, R., Goicoechea, L.~J., Shalyapin, V.~N., \& Oscoz, A.\ 2018, \aap, 616, A118 
\bibitem[Goicoechea \& Shalyapin(2016)]{2016A&A...596A..77G}
Goicoechea, L.~J., \& Shalyapin, V.~N.\ 2016, \aap, 596, A77 
\bibitem[Gordon et al.(2003)]{2003ApJ...594..279G} 
Gordon, K.~D., Clayton, G.~C., Misselt, K.~A., Landolt, A.~U., \& Wolff, M.~J.\ 2003, \apj, 594, 279 
\bibitem[Hainline et al.(2013)]{2013ApJ...774...69H} 
Hainline, L.~J., Morgan, C.~W., MacLeod, C.~L., et al.\ 2013, \apj, 774, 69
\bibitem[Hamuy et al.(1992)]{1992PASP..104..533H} 
Hamuy, M., Suntzeff, N.~B., Heathcote, S.~R., et al.\ 1992, \pasp, 104, 533 
\bibitem[Hamuy et al.(1994)]{1994PASP..106..566H} 
Hamuy, M., Suntzeff, N.~B., Heathcote, S.~R., et al.\ 1994, \pasp, 106, 566 
\bibitem[Horne(1986)]{1986PASP...98..609H} 
Horne, K.\ 1986, \pasp, 98, 609 
\bibitem[Howell(2006)]{howell06} 
Howell, S.~B.\ 2006, Handbook of CCD Astronomy (Cambridge, Cambridge Univ. Press)
\bibitem[Keeton(2001)]{2001astro.ph..2340K} 
Keeton, C.~R.\ 2001, eprint arXiv:astro-ph/0102340
\bibitem[Keeton(2010)]{2010GReGr..42.2151K} 
Keeton, C.~R.\ 2010, GReGr, 42, 2151 
\bibitem[Kochanek(2004)]{2004ApJ...605...58K} 
Kochanek, C.~S.\ 2004, \apj, 605, 58 
\bibitem[Komatsu et al. (2009)]{2009ApJS..180..330K} 
Komatsu, E., Dunkley, J., Nolta, M.~R., et al.\ 2009, \apjs, 180, 330
\bibitem[Kostrzewa-Rutkowska et al.(2018)]{2018MNRAS.476..663K} 
Kostrzewa-Rutkowska, Z., Kos\l{}owski, S., Lemon, C., et al.\ 2018, \mnras, 476, 663 
\bibitem[Koyamada et al.(2017)]{2017ApJ...851...88K} 
Koyamada, S., Misawa, T., Inada, N., et al.\ 2017, \apj, 851, A88 
\bibitem[Krogager(2018)]{2018arXiv180301187K} 
Krogager, J.~K.\ 2018, arXiv:1803.01187
\bibitem[Krogager et al.(2018)]{2018A&A...619A.142K} 
Krogager, J.~K., Noterdaeme, P., O'Meara, J.~M., et al.\ 2018, \aap, 619, A142
\bibitem[Ledoux et al.(2015)]{2015A&A...580A...8L} 
Ledoux, C., Noterdaeme, P., Petitjean, P., \& Srianand, R.\ 2015, \aap, 580, A8 
\bibitem[Lemon et al.(2018)]{2018MNRAS.479.5060L} 
Lemon, C.~A., Auger, M.~W., McMahon, R.~G., \& Ostrovski, F.\ 2018, \mnras, 479, 5060 
\bibitem[Liszt(2014)]{2014ApJ...783...17L} 
Liszt, H.\ 2014, \apj, 783, A17
\bibitem[Lusso et al.(2018)]{2018ApJ...860...41L} 
Lusso, E., Fumagalli, M., Rafelski, M., et al.\ 2018, \apj, 860, A41 
\bibitem[Ma et al.(2017)]{2017MNRAS.472.2196M} 
Ma, J., Ge, J., Zhao, Y., et al.\ 2017, \mnras, 472, 2196
\bibitem[Ma et al.(2018)]{2018MNRAS.474.4870M} 
Ma, J., Ge, J., Prochaska, J.~X., et al.\ 2018, \mnras, 474, 4870
\bibitem[McLeod et al.(1998)]{1998AJ....115.1377M} McLeod, B.~A., Bernstein, G.~M., Rieke, 
M.~J., \& Weedman, D.~W.\ 1998, \aj, 115, 1377
\bibitem[Mediavilla et al.(2005)]{2005ApJ...619..749M} 
Mediavilla, E., Mu{\~n}oz, J.~A., Kochanek, C.~S., et al.\ 2005, \apj, 619, 749
\bibitem[Mediavilla et al.(2011)]{2011ApJ...730...16M} 
Mediavilla, E., Mu\~noz, J.~A., Kochanek, C.~S., et al.\ 2011, \apj, 730, 16
\bibitem[M\'enard \& Chelouche(2009)]{2009MNRAS.393..808M} 
M\'enard, B., \& Chelouche, D.\ 2009, \mnras, 393, 808
\bibitem[M{\o}ller et al.(2013)]{2013MNRAS.430.2680M} 
M{\o}ller, P., Fynbo, J.~P.~U., Ledoux, C., \& Nilsson, K.~K.\ 2013, \mnras, 430, 2680
\bibitem[More et al.(2016)]{2016MNRAS.456.1595M}
More, A., Oguri, M., Kayo, I., et al.\ 2016, \mnras, 456, 1595 
\bibitem[Oke(1990)]{1990AJ.....99.1621O} 
Oke, J.~B.\ 1990, \aj, 99, 1621
\bibitem[{\O}stman et al.(2008)]{2008A&A...485..403O} 
{\O}stman, L., Goobar, A., \& M\"{o}rtsell, E.\ 2008, \aap, 485, 403
\bibitem[P\^aris et al.(2014)]{2014A&A...563A..54P} 
P\^aris, I., Petitjean, P., Aubourg, \'E., et al.\ 2014, \aap, 563, A54
\bibitem[P\^aris et al.(2018)]{2018A&A...613A..51P} 
P\^aris, I., Petitjean, P., Aubourg, \'E., et al.\ 2018, \aap, 613, A51
\bibitem[Pelt et al.(1996)]{1996A&A...305...97P} 
Pelt, J., Kayser, R., Refsdal, S., \& Schramm, T.\ 1996, \aap, 305, 97
\bibitem[Prochaska et al.(2017)]{prochaska17} 
Prochaska, J.~X., Tejos, N., Crighton, N., et al.\ 2017, linetools/linetools:
Third Minor Release, Zenodo, doi:10.5281/zenodo.1036773
\bibitem[Rubin et al.(2018)]{2018ApJ...859..146R} 
Rubin, K.~H.~R., O'Meara, J.~M., Cooksey, K.~L., et al.\ 2018, \apj, 859, A146 
\bibitem[Schneider et al.(2006)]{2006glsw.conf.....M} 
Schneider, P., Kochanek, C.~S., \& Wambsganss, J.\ 2006, Gravitational Lensing: Strong, Weak \& Micro, 
Proc. of the 33rd Saas-Fee Advanced Course, ed. G. Meylan, P. Jetzer, \& P. North (Berlin, Springer)  
\bibitem[Sergeyev et al.(2016)]{2016MNRAS.456.1948S}
Sergeyev, A.~V., Zheleznyak, A.~P., Shalyapin, V.~N., \& Goicoechea, L.~J.\ 2016, \mnras, 456, 1948  
\bibitem[Shalyapin \& Goicoechea(2017)]{2017ApJ...836...14S}
Shalyapin, V.~N., \& Goicoechea, L.~J.\ 2017, \apj, 836, A14 
\bibitem[Sluse et al.(2007)]{2007A&A...468..885S} 
Sluse, D., Claeskens, J.~F., Hutsem\'ekers, D., \& Surdej, J.\ 2007, \aap, 468, 885
\bibitem[Sluse et al.(2011)]{2011A&A...528A.100S} 
Sluse, D., Schmidt, R., Courbin, F., et al.\ 2011, \aap, 528, A100
\bibitem[Smette et al.(1992)]{1992ApJ...389...39S} 
Smette, A., Surdej, J., Shaver, P.~A., et al.\ 1992, \apj, 389, 39 
\bibitem[Tewes et al.(2013)]{2013A&A...553A.120T} 
Tewes, M., Courbin, F., \& Meylan, G.\ 2013, \aap, 553, 120 
\bibitem[Tie \& Kochanek(2018)]{2018MNRAS.473...80T}
Tie, S.~S., \& Kochanek, C.~S.\ 2018, \mnras, 473, 80
\bibitem[Treu(2010)]{2010ARA&A..48...87T} 
Treu, T.\ 2010, \araa, 48, 87 
\bibitem[Ull\'an et al.(2006)]{2006A&A...452...25U} 
Ull\'an, A., Goicoechea, L.~J., Zheleznyak, A.~P., et al.\ 2006, \aap, 452, 25 
\bibitem[Vladilo et al.(2008)]{2008A&A...478..701V}
Vladilo, G., Prochaska, J.~X., \& Wolfe, A.~M.\ 2008, \aap, 478, 701  
\bibitem[Vuissoz et al.(2008)]{2008A&A...488..481V} 
Vuissoz, C., Courbin, F., Sluse, D., et al.\ 2008, \aap, 488, 481 
\bibitem[Wolfe et al.(1986)]{1986ApJS...61..249W} 
Wolfe, A.~M., Turnshek, D.~A., Smith, H.~E., \& Cohen, R.~D.\ 1986, \apjs, 61, 249
\bibitem[Wucknitz et al.(2003)]{2003A&A...405..445W} 
Wucknitz, O., Wisotzki, L., L\'opez, S., \& Gregg, M.~D.\ 2003, \aap, 405, 445
\bibitem[York et al.(2000)]{2000AJ....120.1579Y} 
York, D.~G., Adelman, J., Anderson, J.~E., et al.\ 2000, \aj, 120, 1579
\bibitem[Zuo et al.(1997)]{1997ApJ...477..568Z} 
Zuo, L., Beaver, E.~A., Burbidge, E.~M., et al.\ 1997, \apj, 477, 568 

\end{thebibliography}
\end{document}